\documentclass[preprint2]{aastex} 

\shorttitle{A possible mechanism of origin of heavy elements}
\shortauthors{Tito and Pavlov}

\usepackage{graphicx}
\usepackage{epstopdf}
\usepackage{dcolumn}
\usepackage{bm}

\begin{document}

\title{A possible mechanism of origin of heavy elements \\ in the solar system}

\author{E. P. Tito}
\affil{Scientific Advisory Group, Pasadena, California, USA}

\author{V. I. Pavlov} 
\affil{UFR des Math\'{e}matiques Pures et Appliqu\'{e}es-- LML CNRS UMR 8107, \\ Universit\'{e} de Lille 1, 59655
Villeneuve d'Ascq, France \\ 
\vspace{0.4cm}  
\textnormal{November 16, 2013}}

\begin{abstract}
We advance a hypothesis that a collision of a neutron-rich compact object (NRCO) with a massive dense object of the early solar system was responsible for the heavy element enrichment of the system and for the formation of the terrestrial planets.
\end{abstract}


\maketitle

\section{Introduction} \label{S:Introduction}

\subsection{Motivation} \label{ss:puzzles}

The mechanism of emersion of the Earth's gold and generally the heavy (post-$Fe$) elements of the solar system, remains  an open question despite popular acceptance of the supernova enrichment hypothesis, which posits that the elements were injected into the solar system by one or multiple nearby supernova explosions. 
Numerous individual 
characteristics of the solar system 
when viewed collectively reveal 
that the supernova enrichment scenario is not sufficiently self-consistent. 
In this paper, we suggest an alternative mechanism  
which may provide a better insight into the chemical and physical evolution of the solar system. 
Before laying out the proposed mechanism, 
we list some of these 
characteristics 
and their current explanations.

{\underline {Cosmo-chemical characteristics:}} 

(1)
{\em Presence of stable $r$- and $s$-process elements. }
It is  established that elements beyond $Fe$ are produced via neutron capture by seed nuclei 
only if both abundant free neutrons and heavy nuclei are simultaneously available for the reactions to proceed.
Because the half--life of free neutrons is only  $\sim 15$ minutes,
either the entire episode of heavy elements formation must be of short duration,
or the free neutrons must continuously become available.
In nature, such environments are known to exist either during the collisions of neutron stars,
or in the interiors of giant stars, in which case the only way for the elements to be released is by 
the star 
explosions.
Thus, currently it is assumed that those solar system elements that are 
theoretically produced only by the {\em rapid ($r$-)} and/or {\em slow ($s$-)} processes, 
 were actually produced in explosive stellar events and
delivered to our system by propagating shock waves and winds.

(2)
{\em Presence of short-lived nuclides.}  
There is abundant evidence that short-lived nuclides
once existed in meteorites. 
On a galactic scale, red giants and supernovae continually inject newly synthesized elements into the interstellar medium,
and unstable nuclides steadily decay away.
These two competing processes result in steady-state abundance of these nuclides in the interstellar medium.
The abundances of some of such discovered nuclides ($^{107}Pd$, $^{129}I$, $^{182}Hf$, for example) roughly match the expected steady-state
galactic abundances and hence do not necessarily require a specific synthesis event.
However, the appearance of 
$^{26}Al$, $^{41}Ca$, $^{53}Mn$, $^{60}Fe$, and a few other nuclides, in the early solar system require synthesis of them 
at the time, 
or just before, the solar system formed.%
\footnote{
See, among others, reviews by 
\cite{w06}, \cite{w07}, 
and references therein.
} 

The conventional view is that these nuclides were synthesized in a nearby supernova and/or a red 
giant  
and injected into the solar nebula just shortly before the solar system formation.%
\footnote{
See, among others, 
\cite{ct77}, \cite{c95}, \cite{bf98}, \cite{gv00}, \cite{l06}, 
and references therein.
} 
However, 
various numerical 
models of stellar nucleosynthesis 
repeatedly 
show that 
one 
event by itself 
cannot provide the early solar system with the full
inventory of short-lived nuclides.   
Depending on the model, certain isotopes are significantly  over- or under-produced.%
\footnote{ 
See, among others, 
\cite{g09}, \cite{g12}, 
and references therein.%
} %
%
%
Meteoritic sample studies concur by revealing data signatures inconsistent with a single stellar origin.   
For example, 
 the Ivuna CI chondrite analysis 
detected 
simultaneous presence of at least 
5 mineralogically distinct carrier phases 
for $Mg$ and $Ca$ isotope anomalies,  
leading to the explanation that they must represent 
"the chemical memory of {\em multiple and distinct} stellar sources"   
\citep{s12}. 

(3) 
{\em Challenges to  Supernova Hypothesis.} 
Besides its inability to explain with one event the entire inventory of the enrichment elements, 
the supernova hypothesis faces additional challenges.

On the one hand, 
to be able to provide the observed abundances of radioactive isotopes,  
the supernova must have been located not too far from the solar nebula ($d < d_{max}$). 
On the other hand, the distance had to be great enough ($d > d_{min}$) so that the shock wave from the supernova 
did not destroy the nebula.  
 For  the 
 stars with $M \sim 25 M_{Sun}$   
 shown to 
provide the best ensemble of short-lived radioactive nuclei,  
this optimal range is quite narrow, $d \simeq 0.1-0.3$~pc \citep{a10}. 

Furthermore, 
stars within the cluster typically form within 1-2~Myr 
and  the clusters disperse in about 10~Myr or less. 
Since stars with mass $M \sim 25 M_{Sun}$ burn for $\sim 7.5$~Myr before core collapse 
\citep{w02}, 
to fit the supernova enrichment scenario 
the Sun must have formed several Myr after the progenitor  
\citep{a10}. 
If located $\sim 0.2$~pc  from the progenitor, 
the early solar nebula could have been evaporated by the progenitor radiation. 
One way to reconcile this is to assume that the early solar nebula 
and the progenitor approached each other at the $0.2$~pc separation distance just before the supernova explosion  \citep{a10}.
Such timing requirement lowers the odds for the supernova enrichment theory.

Moreover, if the above-mentioned radionuclides were produced by {\em multiple} stellar sources, 
all of these injection events, as well as the subsequent highly homogeneous mixing of isotopes,  
had to occur within the time-span of only about 20,000 years,
as constrained by the spread of calcium-aluminum inclusions (CAI's) condensation ages. \citep{g12}

(4)  
{\em Presence of $^{10}Be$ and $^{7}Li$ 
isotopes. }
Detection of $^{10}Be$ indicates
that  
one more process, {\em local} to the solar system, must be added to the enrichment scenario. 
$^{10}Be$  is not synthesized in stars.  Indeed, in most stellar events $Be$ is destroyed rather than produced.
Moreover, the discovered excess of $^{7}Li$ in CAI 
(\cite{c01}; \cite{crm02}) 
points with certainty to its origin {\em within} the solar system,
because $^{7}Li$ is produced by decay of $^{7}Be$ whose half-life is only 
53~days.
It was suggested that these elements were produced by spallation within the solar system as it was forming.
Various groups tested this
scenario by comparing the modeled nuclear spallation yields with the inferred solar system initial ratios 
(e.g., \cite{l98}; \cite{g01}; \cite{gms01};  \cite{l03}). 
However, they 
failed to 
self-consistently explain the abundance discrepancies.

 (5)
 {\em Presence of $p$-process nuclides. }
 A number of proton-rich isotopes ($p$-nuclei) existing in the solar system,  
cannot be made in either the $s$- or the $r$-process.
  Although their solar system abundances are tiny compared with
isotopes produced in neutron-capture nucleosynthesis, the site of their
production in the solar system is even more problematic.
They can be produced either by the proton-capture from elements with lower charge number, or by photodisintegrations.
Both production mechanisms require high temperatures and presence of seeds ($s$- and/or $r$-process nuclides).
Proton capture process also requires a very proton-abundant environment.

 Currently, the solar system abundances of $p$-nuclei have been best fitted into the combination of contributions from {\em several} stellar processes.
 Photodisintegration in massive stars (Type Ia supernova or a mass-accreting white dwarf explosion; see \cite{r13}) and
  neutrino processes (for $^{138}La$  and $^{180}Ta$), can perhaps explain the bulk of the $p$-nuclei abundances.
  However,
the abundances of light $p$-nuclei in the solar system  significantly exceed  the simulated production 
from the stellar processes, and this 
problem 
has not yet been resolved 
\citep{r13}.

{\underline {Planetary structure characteristics:}}

Besides the above-mentioned peculiarities of its chemical composition, 
 the solar system exhibits some unusual characteristics in its planetary structure as well.

(6)
{\em Orbits of giants. }
Unlike the bulk of known exoplanetary systems, the orbits of the solar system's giant planets are 
remarkably 
widely spaced and nearly circular.
(See, for example, overviews in \citet{f-a01} and \citet{b-a04}). 
$N$-body studies of planetary formation  and orbit positions indicate that, due to the convergent planetary migration in times before the gas disk's dispersal, each giant planet should have become trapped in a resonance with its neighbor \citep{k00, ms01}. 
To explain its present, stretched and relaxed state,  an evolution scenario is required where the outer solar system underwent a violent phase when planets scattered off of each other and acquired eccentric orbits \citep{t-a99, t-a05},
followed by the subsequent stabilization phase.

(7)
{\em Two classes of planets. }
The solar system also features two distinct types of planets: the inner terrestrial and the outer giant (jovian) ones.  The rocky terrestrial planets are thought to be formed by accretion (from dust grains into larger and larger bodies). However, there is some uncertainty in the understanding as to how the giant  planets formed.  There are essentially two classes of theories.  The earlier one is that of \emph{core accretion}.  It proposes that rocky, icy cores of giant planets accreted in a process very similar to the one that formed the terrestrial planets and then captured gas from the solar nebula to become gas giants. This theory explains why the  giants have larger concentration of heavier elements than the Sun, but unfortunately numerical simulations yield formation times that are way too long unless the mass of the primordial nebula is increased. The second theory posits that a density perturbation in the disk could cause a clump of gas to become massive enough to be self--gravitating and form the Sun and the planets \citep{b97}. 
Formation scale is then much more rapid, but the theory does not readily explain the observed chemical enrichment of the planets.

(8)
{\em A missing  giant. }
There are also indications that one more giant object initially may have been present in the solar system, but  
somehow disappeared 
at some point. 
For example, \citet{n11} 
attempted to determine which initial states were plausible and the findings showed that dynamical simulations starting with a resonant system of four giant planets had low success rate in matching the present orbits of giant planets combined with other constraints (e.g., survival of the terrestrial planets).
A fifth giant had to be assumed to produce reasonable results.

\subsection{Hypothesis}

We 
advance a hypothesis which is capable of  a self-consistent reconciliation of all of the above-mentioned peculiarities by envisioning only {\em one} event. 

We
suggest that our solar system initially 
possessed \emph{no terrestrial} but \emph{only jovian} planets.
Possibly, it possessed another massive dense object closest to the Sun (see discussion in Section~\ref{ss:target}).
We further propose that close to five billion years ago 
(i.e., at the currently assumed birth time of the solar system, which is based on meteorite age estimates) 
either this additional object, or the edge of the Sun if no extra object existed, 
was hit by 
a fast-moving  
 {\em neutron-rich compact object} (NRCO) resembling a compact neutron star.
As a result of this collision, 
the interior matter of the NRCO  
was ejected (see further discussion) into the 
solar system. 
A multitude of nuclear reactions and transformations (see further discussion) followed:  
fragmentation of the "neutron droplets", fission of giant nuclei, 
formation of abundant free neutrons and protons, 
production of 
$r$-, 
$s$-, 
and $p$-process
elements, including the above-mentioned short--lived radionuclides, isotopes, and so on. 
The resulting products 
of this transformation chain  eventually formed the modern terrestrial planets and other rocky bodies,
and also enriched the pre-existing 
jovian planets 
with 
additional 
chemical elements.

Indeed, 
collisions of neutron stars (in black-hole/ neutron star or two neutron star mergers) 
  have been considered (see, among others, \cite{ls76, f-a99}). 
A chance encounter of the solar system with another star has also been suggested, in an attempt to provide a plausible explanation for the orbit of Sedna 
\citep{kb04}.    
But the idea of a direct 
collision of a neutron-rich object with the solar system has never been advanced.%
\footnote{
In general, exotic stellar collisions involving planetary systems and neutron stars have been 
contemplated.  
For example, \cite{s03} proposed explaining the origin of a triple "neutron star - white dwarf - planet" system 
detected in the distant globular cluster M4, 
by a collision and exchange 
of a main-sequence star system with a neutron star binary.} %

We suspect 
that the proposed collision  is a unique and extraordinary event. 
However, despite its small statistical odds, 
the collision hypothesis has the advantage over other currently considered theories 
by self-consistently explaining  with only one event  the production of {\em every} chemical element ever detected in the solar system.  
 The planetary structure of the solar system (with two classes of planets and atypically wide, circular and non-resonant orbits of giants)
  also fits well within the collision scenario. 
  Finally, the collision hypothesis does not contradict {\em any} of the already developed theories, 
  but rather integrates them all from a different perspective. 
  All of the existing  enrichment models are valid conceptually, 
  but the relative contributions from various mechanisms perhaps are less than they have been assumed so far.

The 
remainder of the 
article is structured as follows:
Section~\ref{S:object} considers the properties of the colliding object.  
Section~\ref{S:collision} outlines the collision process. 
Section~\ref{S:conclusion} discusses and summarizes the results.
Appendices~A-D present the details.

\section{Colliding object}  \label{S:object}

\subsection{Modern neutron stars} \label{ss:NS}

Modern neutron stars are the closest "role models" for the neutron-rich compact object (NRCO)
that delivered the abundant 
neutrons into the solar system according to the outline scenario.

Current understanding is that neutron stars are compact, gravitationally-powerful stars composed of strongly degenerate,
predominantly neutron matter (with an admixture of super-conducting protons, electrons, and muons).%
\footnote{
\cite{p10}; \cite{st83}; \cite{zn96}; \cite{g00}; \cite{hpy07}; \cite{yls99}; \cite{b99}.
} 
At present, neutron stars are thought to have typical masses $M \sim 1.4 M_{sun}$ and radii $R \sim 10 \, km$ 
(which 
are not directly observable, 
but inferred from model calculations of X-ray bursters).  
Neutron star temperatures are thought to be of order of $10^{11} K$ immediately after formation,
falling rapidly thereafter to $\sim 10^6 K$ (crust) $- 10^8 K$ (core).%
\footnote{
$10^{11} K \simeq 10 MeV$ 
which is 
in the range of estimates and measurements for   the critical temperature of 
nuclear gas-liquid transition in a nucleus, $T_c \sim 5-18 MeV$ \citep{k09}.
} 
The outer regions of neutron stars are thought to be solidified and form a $Fe$-rich crust (with thickness of $\sim 1 \, km \ll 10 \, km$ of radius).
Once the inward--increasing density reaches $\rho_0 \approx  2.8 \times 10^{14} g / cm^3$,
the theory predicts that
the nuclei of the matter become so close to each other that they merge to form nuclear liquid.
In such a high density environment 
(above $\rho_{drip} \simeq  4 \times 10^{11} g / cm^3$) 
the $r$-process and beta-decay process are suppressed,
and no individual complex nuclei form.
Thus, the core of a neutron star represents effectively one "giant nucleus" composed predominantly of neutrons.

The equation of state in a neutron star {\em crust} has been calculated with an accuracy sufficient to construct neutron star models,
although some theoretical problems still remain unsolved.
By contrast, the equation of state in the {\em interior} where $\rho \geq \rho_0$ cannot be calculated exactly
because of the lack of the precise relativistic many--body theory of strongly interacting particles. 
Instead of the exact theory, there are many theoretical models whose reliability decreases with growing $\rho$. \citep{ls91}. 
Thus, the equation of state in neutron star cores remains largely unknown
\citep{hpy07, b91, pr91}.
Appendix \ref{A:NSmodels} lists some of the models of equations of state for nuclear matter and neutron stars.

Despite being composed of $Fe$-lattice, the crust of a neutron star can crack. 
Several causes and mechanisms responsible for the cracking of the crust of neutron stars have been considered.
In neutron stars with strong magnetic fields (magnetars),
the stress of the twisting force of the magnetic field can reach the levels at which the crust breaks. 
In fast-rotating neutron stars (pulsars), when their spins slow down rapidly, 
the crust cracks from within
because the "fluid" inner matter forming the rotational bulge redistributes and exerts stress on the solid crust.
When neutron stars are subjected to a powerful gravitational field
(for example in the proximity of a black hole or in  a tidal lock-up with another star),
the crust cracks when its breaking strain is exceeded due to the 
tidal deformation.  
Neutron star crust is 
highly conductive, thus a 
magnetic field has strong influence on the rupture dynamics -- 
sometimes the crust shatters suddenly, sometimes the crack is simply a propagating rupture. 
\citep{hk09, ch10, t11}

\subsection{Neutron-Rich Compact Object (NRCO)} \label{ss:NRCO}

The above-mentioned "traditional" neutron stars are not the only potential sources of neutrons that have been theoretically considered.
Indeed,  a number of exotic compact stars have been hypothesized,     
such as:  "quark stars" --  a hypothetical type of stars composed of quark matter, or strange matter;
"electro-weak stars" -- a hypothetical type of extremely heavy stars,
in which the quarks are converted to leptons through the electro-weak interaction,
but the gravitational collapse of the star is prevented by radiation pressure;
"preon stars" -- a hypothetical type of stars composed of preon matter.  
Thus, various neutron-rich objects can exist or could have existed 5~Gyrs ago.

In the proposed collision scenario, the NRCO must  possess 
only one main 
characteristic: 
be able to deliver to the solar system abundant, over-saturated with neutrons, hyper-nuclei  
that are critical for the formation of the above-mentioned chemical elements.  
We also assume that the NRCO is gravitationally-powerful, 
and therefore  is compact and its interior neutron matter is highly dense, forming neutron liquid inside and $Fe$-rich crust outside, 
similarly to the traditional neutron stars. 
Other characteristics of the NRCO (for example, its size or equation of state) may differ from the commonly used models of neutron stars,   
 as long as the existence of such neutron-rich object is theoretically permitted (see Appendix~\ref{A:NRCO}).

\section{The Collision Scenario} \label{S:collision}

\subsection{Ejecta and Energy Considerations} \label{ss:energy}

A critical question for our scenario is whether a portion of the NRCO's mass can 
be permanently ejected into the solar system.%
\footnote{
To fully tear the NRCO apart into individual nucleons, 
the energy of order  $Q \sim  G M^2 / R$ (gravity potential energy, not including energy of other mechanisms) is needed.  
Per nucleon, it is 
$Q_1 \sim G M^2 / R N_n$ = $ G m_n^2  N_n  / R$ = $m_n c^2 (G M /c^2) / R$ = $m_n c^2 (R_g / R)$ = $(R_g / R) \times 1 \, GeV$.  
Even when $R_g / R$ (which depends on the NRCO mass) is a small fraction of 1,  
this energy amount is substantial. 
(However, to split the NRCO into two pieces, rather than completely apart into individual nucleon, 
much less energy is needed.)
} %
%

Under most initial conditions, the  change of the NRCO's kinetic energy as a result of collision 
is  likely to be insufficient 
for the ejecta to overcome the NRCO's gravitational pull.%
\footnote{
The total mass of all four terrestrial planets is $\sim 2 \, M_{Earth}$.  
But the interiors of the jovian planets may contain some of the hypothesized ejecta. 
Jupiter, for example, is thought to possess a solid core  $\sim 15 \, M_{Earth}$,  
however, not all of it had to originate from the ejecta. 
A simple non-relativistic calculation shows that 
the amount of energy needed 
to permanently eject $10 \, M_{Earth}$ into the solar system
is comparable with the kinetic energy released when 
a NRCO with, for example, $M = M_{Sun}$, $R = 10 \, km$, and $V_{beg} \sim 1,000 \, km/s$ 
slows down to $V_{end} \sim 450 \, km/s$. 
} %
%
 However, theoretical and experimental results from the high-energy nuclear physics (see Appendix~\ref{tc})   
 establish that reactions of multi-fragmentation of nuclear hyper-nuclei (as well as fission of super-nuclei)  
 are capable of releasing  amounts of energy that are  
 comparable and even exceeding the levels necessary for the permanent ejecta.  
 In the core of the NRCO, such reactions would occur 
 if  density of the nuclear matter (in a localized spot) drops below $\rho_{drip} \simeq 4 \times 10^{11} \, g/cm^3$.  
 (In fact, as discussed in Section~\ref{ss:explosion} and Appendix~\ref{tc}, 
 a number of nuclear reactions would then occur: $\beta$-decay, $\gamma$-emission, multi-fragmentation, fission, etc.)
 
 The mechanism that can produce such decompression (from $\rho \sim 10^{14} \, g/cm^3$ to $\rho \sim 10^{11} \, g/cm^3$) 
 is the {\em sufficiently strong} deceleration of the NRCO during the collision.  
 In the "head-on" collision,  redistribution of core mass due to deceleration compresses the front part of the core and decompresses the rear. 
 Notably, only a  small region of the overall core needs to decompress to below $\rho_{drip}$ 
 in order to to create a sufficiently powerful explosion.%
 \footnote{ 
The corresponding increase of density in the compressing part is therefore small.} %

 Thermodynamically, decompression from $\rho \sim 10^{14} \, g/cm^3$ to $\rho \sim 10^{11} \, g/cm^3$ occurs in two steps.  
 First, 
 the density drops from its initial state to the phase-transition boundary, 
 which separates stable "nuclear liquid" from the unstable "nuclear fog"  
 (discussed in detail in Section~\ref{ss:explosion} and Appendix~\ref{tc}).  
 Second, once the matter is in the state of the "nuclear fog", the instability of the (negative) density perturbation develops exponentially fast, 
 leading to rapid and substantial further density decrease by a factor of $10^2$ or more.

 Therefore, it is very important to note that 
 in the context of this analysis, 
 the strength 
 of the NRCO's deceleration is defined 
 not in the kinetic energy sense, 
 but in the phase-transition sense.   
 The {\em sufficiently} strong deceleration 
 is such deceleration that is capable of 
 producing enough density decompression  in the rear of the NRCO 
 (by redistributing the core mass within the solid crust shell away from the back, relative to the flight path) 
 that the initial thermodynamical phase state (nuclear liquid) of some rarified localized increment in the back,  
 shifts to the phase boundary, and crosses it, into the unstable two-phase domain  of "nuclear fog".

Obviously, the closer the initial phase state of the NRCO's core matter is to the phase boundary, 
the smaller the instant deceleration magnitude needs to be to produce the shift. 
Thus, in certain circumstances, the magnitude of deceleration that may appear to be small in the kinetic energy sense, 
may be sufficiently large in the phase-transition sense.

Unfortunately, 
as stated in Section~\ref{ss:NS}, 
the equation of state (EoS) in the neutron star interior remains uncertain 
because of the lack of the precise relativistic many-body theory of strongly interacting particles.  
Moreover, 
even the precise EoS would not accurately represent the evolution of the system 
that is undergoing the above-mentioned phase transition 
and instability development,  
because the EoS is always defined for the system in (quasi)-equilibrium, not a fast-evolving one. 
Therefore, 
at this stage, any 
attempts to quantify in detail the density perturbations 
would not be meaningful.

It is also important to keep in mind that the nuclear reactions occur with time scales ($\sim 10^{-22} \div 10^{-15} \, sec$)  
that are dramatically shorter 
than the time scales of the macro processes involved 
(such as the NRCO flight, or even the core density redistribution.) 
Thus even a very short-lived strong perturbation of density may be able to start the nuclear reaction cascade 
(see Section~\ref{ss:explosion} and Appendix~\ref{tc}).  

Finally, it is worth noting that the above-mentioned density decompression occurs in the proposed collision scenario 
because the NRCO
is decelerating 
{\em longitudinally}.  
Such decompression mechanism does not arise in supernova (because of its radial symmetry), 
nor 
in a spiraling interaction with another stellar object.

\subsection{Collision Target} \label{ss:target}

In view of the discussion above, 
we list the following candidates for the target within the solar system. 

A number of independent analyses have pointed at the potential existence of an additional giant object 
  in the early solar system. 
 (See \cite{n11}, \cite{b11}, \cite{p92}). 
 Thus, one candidate could be a large gaseous planet -- a "super-Jupiter" perhaps -- 
rotating around the  Sun on the first orbit 
(located inside the Jupiter's orbit, which was second at that time).

 Furthermore, not to miss any options, could the Sun have had a close binary companion?  
 If yes, the companion (whether a white  dwarf, or a brown dwarf, or a main-sequence star) could also possibly fit the role of the target.%
 \footnote{
  The majority of solar-type stars are found in binary systems.  (\cite{a83}, \cite{dm91}, 
 \cite{k11}). 
The well-known problems with angular momentum dispersal (e.g., \cite{b95} 
and references therein) indicate that  protostars should end up in binary or multi-stellar formation. 
Furthermore, 
the $7^o$  misalignment between the Sun's 
rotation axis and the north ecliptic pole (see, e.g., \cite{bg05}), 
may indeed be supportive of such scenario.
In our case,  both companions  would have had to form a close binary  
 and remain inside the orbit of Jupiter (wherever it was positioned at that time). 
 A white dwarf would have been an ideal candidate for the target because it is very dense and rather massive, 
 and could likely produce the sufficient deceleration of the NRCO.  
 An average white dwarf has mass $\sim 0.5-0.6 M_{Sun}$, density $\sim 10^6 \, g/cm^3$ and size $\sim  R_{Earth}$. 
}%

Finally, a scenario can be envisioned   in which 
 the NRCO flies through the edge of the Sun (without destroying it), 
significantly decelerates (in the sense defined above), 
undergoes the above-mentioned decompression and nuclear reactions in its rear (as mentioned above and discussed in detail in the subsequent sections), 
and ejects a portion of its inner matter post-collision, at a distance $\sim 1 A. U.$. 
In this scenario, the target is the Sun.  
No additional solar system object is then required to have existed, but the general hypothesis of element formation could still be valid. 

For all of these alternatives, the key question is whether the NRCO can sufficiently decelerate upon the collision 
for the decompression and the subsequent nuclear reactions to take place.

\subsection{Deceleration} \label{ss:deceleration}

As the NRCO penetrates the target, 
it decelerates 
due to 
a number of
effects, such as 
classical drag, 
 dynamical friction, 
 acquisition of target particles onto the gravitationally-powerful NRCO, 
Cherenkov--like radiation of various waves related to collective motions generated within the target, 
distortion of the magnetic fields,
and possibly others.
Obviously, some deceleration causes are dominant and some are negligible.

Unfortunately, as stated earlier, since the strength of deceleration is defined in the context of nuclear phase transition (Section~\ref{ss:energy}),  
but the equation of state for the NRCO  or neutron stars is significantly uncertain (Section~\ref{ss:NS}), 
no  quantitative analysis of deceleration mechanisms would be meaningful at this point.  
(However, this should not prevent the formulation of the hypothesis in general.)
Nonetheless, we demonstrate what the NRCO's deceleration may look 
like (in the kinetic sense) 
for several sets of initial assumptions.    

Considering how gravitationally powerful the NRCO is, 
for simplicity we consider only one of the deceleration mechanisms --  acquisition of the particles of the target. 
Because the effect is more pronounced when the target is dense and massive, 
we developed a model specifically for such target.  
The effect of accretion on neutron stars has been studied in a variety of settings,  
for example, for particles from interstellar medium, a stellar wind, a common envelope in a binary system, or a supernova ejecta 
(see Appendix~\ref{A:accretion} and references therein).
However, deceleration of a neutron star due to capture of particles of a very dense, fully-degenerated target (such as a white dwarf, for example)
has never been developed in literature.
Appendix~\ref{A:accretion} presents our model in detail.

Our model is composed of a system of nonlinear equations
describing how speed $\mu (t) $, deceleration $\mu ' (t)$ and mass $m(t)$
of the NRCO change as it transits through the target and captures the surrounding particles along the way.
(See  Eqs.~(\ref{b2d}), (\ref{em1}) and (\ref{em3}) in Appendix~\ref{A:accretion}.)
The equation of state for the target is derived using the statistical description of a fully-degenerated non-relativistic Fermi gas.

\begin{figure}
\centering
\includegraphics[width=8cm]{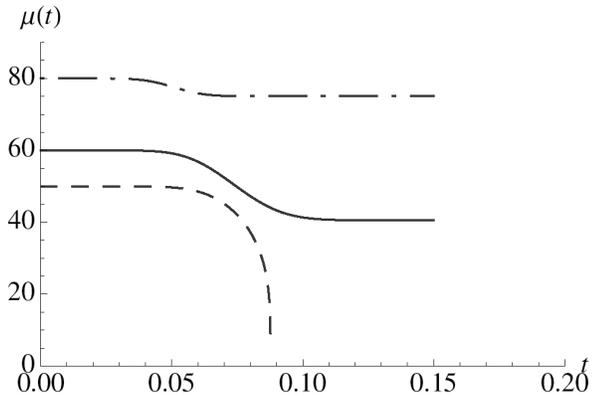} 
\caption{
Evolution of velocity $\mu (t )$ of the colliding NRCO decelerating due to capture of the particles of the target. (Time and velocity are in dimensionless units.) Three initial velocities are considered:
$\mu(0)_A = 60$ units $= 8  \times 10^6 \, m/s$ (solid line),
$\mu(0)_B = 80$ units (dash-dot line), and
$\mu(0)_C = 55$ units (dashed line).}
 \label{DNStarFig1}
\end{figure}

\begin{figure}
\centering
\includegraphics[width=8cm]{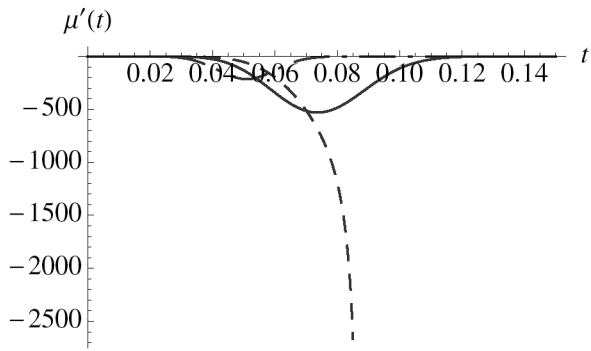} 
\caption{
Evolution of deceleration rate (negative acceleration rate) $\mu ' (t )$ of the colliding NRCO decelerating due to capture of the particles of the target. (Time and deceleration rate are in dimensionless units.)
Three initial velocities are considered:
$\mu(0)_A = 60$ units $= 8  \times 10^6 \, m/s$ (solid line),
$\mu(0)_B = 80$ units (dash-dot line), and
$\mu(0)_C = 55$ units (dashed line).}
 \label{DNStarFig2}
\end{figure}

\begin{figure}
\centering
\includegraphics[width=8cm]{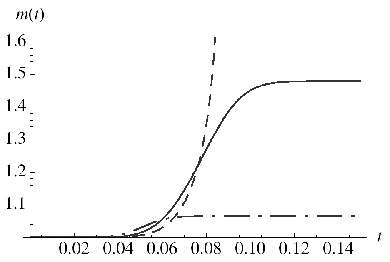} 
\caption{
Evolution of mass $m (t )$ of the colliding NRCO decelerating due to capture of the particles of the target. (Time and mass are in dimensionless units.) Three initial velocities are considered:
$\mu(0)_A = 60$ units $= 8  \times 10^6 \, m/s$ (solid line),
$\mu(0)_B = 80$ units (dash-dot line), and
$\mu(0)_C = 55$ units (dashed line).}
 \label{DNStarFig3}
\end{figure}

Fig.~\ref{DNStarFig1} and Fig.~\ref{DNStarFig2} plot the speed and deceleration of the NRCO as it flies through the target 
(for ease of calculations, we assume it goes through the center), 
while Fig.~\ref{DNStarFig3} shows how the NRCO acquires additional mass during the process.
For illustration purposes we considered
the following initial conditions:
the NRCO's initial mass -- $0.5 M_{Sun}$,
target density -- $\rho = 1.5 \times 10^6 \rho_{Sun}$,
target radius -- $L_n = 0.01 \times R_{Sun} \sim 7 \times 10^3 \, km$.
Time is measured in dimensionless units of $\tau = L_n / (c_1)_c $,
where $L_n$ is target radius and $(c_1)_c$ is the speed of sound in the center of the 
target.
Thus, $\tau \sim 53 \, sec$, and the entire event takes just a few seconds.
With respect to the initial speed of the NRCO $\mu(0)$, we present three scenarios: a base case with $\mu(0)_A = 60$ dimensionless units, or $8 \times 10^3 \, km/s$, and two sensitivity cases with  $\mu(0)_B = 55$ units and $\mu(0)_C = 80$ units. (Many neutron stars have been observed moving with spatial velocities $\sim 10^3 \, km/s$. See for example, \citet{z-a07} and \citet{f96}). 

In the base scenario (case~A, solid line), during its flight, the colliding NRCO added more than 40\% to its mass (Fig.~\ref{DNStarFig3})  and lost about 1/3 of its speed (Fig.~\ref{DNStarFig1}).%
\footnote{
 A good visual analogy for 
 gravitational mass acquisition 
 is a strong magnet hauled through a pile of metal slivers -- once pulled out of the pile, 
 the magnet becomes enormously massive due to magnetic accretion. 
The gravitationally-powerful NRCO captures not only the particles 
 located {\em in front of} it, but also the particles located to the sides. 
 Indeed, one can speculate that 
 the 
 NRCO perhaps does not even need to penetrate the target. It may be able to pass "close enough" and still capture the mass 
 and decelerate. 
 }

Case~B (dash-dot line) corresponds to a higher initial velocity, which means that the duration of the fly-through is shorter and the number of captured particles involved in the process is smaller. This scenario has the least acquisition of mass (Fig.~\ref{DNStarFig3}), the least loss of speed (Fig.~\ref{DNStarFig1}), and the smallest maximum of the magnitude of deceleration (Fig.~\ref{DNStarFig2}).

In Case~C (dashed line) the NRCO completely stops (Fig.~\ref{DNStarFig1}). It is important to keep in mind that such outcome should be considered as an asymptotic scenario. Our model is imprecise in the sense that the Bondi accretion interpolation formula assumes an infinite-sized nebula. Obviously, due to the law of mass conservation the NRCO cannot capture more than the mass of the target. Even if it captures a substantial portion of the target, the model become no longer accurate as it keeps constant certain properties of the target. Nonetheless, Case~C shows that instant deceleration can reach enormous magnitudes (infinity at full stop) when mass acquisition is significant (Fig.~\ref{DNStarFig2}).

An important general observation can be made from these simulations.  
The instantaneous deceleration magnitude can be rather large (even in the kinetic sense) 
when the NRCO can acquire enough (relative to itself) mass from the target.  
The absolute sizes of the colliding objects do not matter, only the relative do. 
Thus any of the above-mentioned target candidates (a planet, a dwarf, or the edge of the Sun) 
may be able to significantly decelerate the NRCO, 
if the NRCO has an appropriate size. 
(In Appendix~\ref{A:NRCO}, we demonstrate that NRCOs of much smaller sizes 
than traditional neutron stars 
can indeed exist with the equation of state derived for this model.)

\subsection{Density stratification} \label{ss:stratification}

Strong deceleration leads to stratification of the NRCO's core.  
Behavior of an elastic body in the frame of reference moving with acceleration/deceleration is analogous to its behavior in a homogeneous gravity field.
This means that density stratification will always take place.
This effect will be significant if the characteristic scale of stratification is much less than the size of the object.
(The characteristic scale here is defined as $s^2/a$ where $s^2$ is square of the sound speed within the elastic body, and $a$ is deceleration magnitude).

In our scenario, significant stratification means $s^2 / a \ll R_{nrco}$, where $R_{nrco}$ is the size of the NRCO.
The magnitude of deceleration, $a$, may be estimated as $a \sim \Delta V / \tau$,
where $\Delta V$ is the change of the NRCO's velocity over the duration of the deceleration process, $\tau$.
If NRCO deceleration is significant then $\Delta V$ is a meaningful fraction of $V$.
The duration of the process (i.e., the duration of the flight through the target), $\tau$, may be estimated as $\sim R_{target} / V$,
where $R_{target}$ is the size of the target.
Thus, the condition for significant stratification  becomes
\begin{eqnarray}
\frac{s^2}{V_{nrco}^2} \ll \frac{R_{nrco}}{R_{target}}
\end{eqnarray}
Since in our case, $R_{nrco} \ll R_{target}$, it necessarily implies that for significant density stratification to take place,
the elasticity of the NRCO matter (characterized by $s^2 = \partial p / \partial \rho$) must become rather small,
i.e. the mono-phase state of the matter has to be near the boundary of its thermodynamical (gas/liquid) equilibrium.

In Appendix~\ref{A:NRCO}, we offer a model which possesses the necessary characteristics
for such phenomenon.
The NRCO's core described by the equation of state (Eq.~\ref{eos:07}) can 
have
quasi-homogeneous
density distribution, except for a narrow region near the edge
where density gradient is negative and extremely steep.

\subsection{Explosion} \label{ss:explosion}

As stated earlier, 
if the equilibrium state of the inner "nuclear liquid" is initially close
to the boundary of the liquid/gas phase transition, 
then decompression by deceleration can shift its phase state from the liquid phase into the two-phase zone of "nuclear fog".
(We describe this process in detail in Appendix~\ref{A:decompression}.)
In the two-phase zone, the matter can exist as a mixture of two phases of nuclear matter -- 
either liquid droplets surrounded by gas of neutrons, 
or homogeneous neutron liquid with neutron-
gas bubbles.
In such state, the matter can reach substantial further rarification, 
reducing density 
by a factor of $10^2$ or more due to hydrodynamic instability.
(See Appendix~\ref{A:decompression}.) %
Below density $\rho_{drip}$, 
beta-decays are no longer Pauli-blocked and significant amounts of energy become released.  
Indeed, in their simulations of $r$-process nucleosynthesis in neutron star mergers, 
\cite{f-a99} 
observed 
that from the level of $\rho_{drip}$, the density dropped remarkably fast.
The material initially cooled down by means of expansion,
but then started to heat up again when the $\beta$-decays set in.

This energy, absorbed by the nuclear "droplets" of the "fog", triggers fragmentation of these supersaturated hyper-nuclei   
(see Appendices~\ref{A:NRCO} and \ref{A:decompression}; see \cite{b83}, \cite{h88}, \cite{l89}).
These reactions, known to release even more energy ($\sim 1 MeV$ per fission nucleon, as known from trans-uranium fission events), 
proceed effectively at the same time as the beta-decay reactions, 
all occurring with a very rapid nuclear time scale ($\sim 10^{-22} \div 10^{-15} \, sec$).  
When perturbations of the equilibrium of a neutron liquid (droplet) 
permit production of charged protons (even in small numbers, and in small localized regions), 
spontaneous fission reactions 
commence.

\begin{figure}
\centering
\includegraphics[width=8cm]{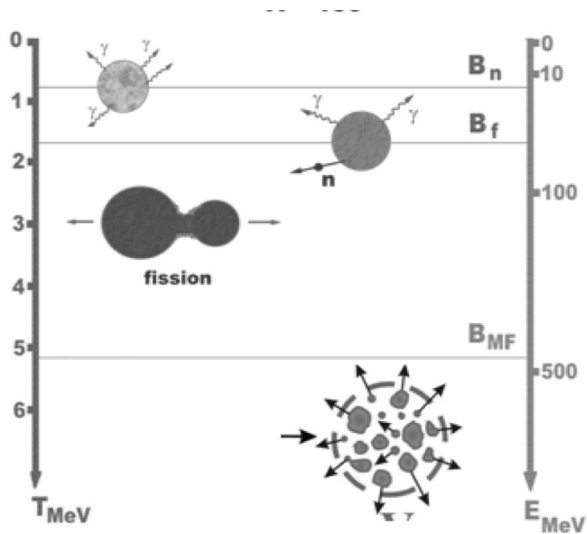} 
\caption{Types of nuclear reactions ($\gamma$-emission, neutron-emission, fission, multi-fragmentation) 
based on the injected energy of excitation.   
From \cite{k-a11}.  
\label{karnaukhov}}
\end{figure}

Fig.~\ref{karnaukhov} (from \cite{k-a11}) schematically explains the process.  
When a nuclei is excited weakly (low $T_{MeV}$), only $\gamma$-emission occurs.  
At a higher level of excitation, neutron-emissions 
start taking place. 
When even  more energy is applied to the nucleus, it deforms and  
fission reactions start, because for deformed nuclei with $Z^2/A > 50$, 
electrostatic repulsion starts exceeding surface tension. 
And finally, when injected energy is sufficiently high, 
splitting into fragments ("droplets" if the initial nucleus is a hyper-nucleus) occurs, 
followed by the 
cascade of further splitting into fragments and neutron emissions.

The energy released in the process can be powerful enough to crack the crust from within and eject 
a portion of the NRCO's inner matter beyond its gravitational trap (see Section~\ref{ss:energy}). 
Once ejected, the neutron "droplets",  hyper-nuclei, and other components of the explosion, 
will transform into a variety of chemical elements.

\subsection{Element production} \label{ss:abundances}

Unfortunately, at this point it is not possible to assess the resulting abundances of the elements produced in the process.

First, the
theory of {\em fission} (and even more so of {\em fragmentation}) of hyper-nuclei ($ln A \gg 1$)
 is not developed at all, mostly because observational data are impossible to collect.
Fission reactions
 (split of elements with high $A$ numbers into several with lower $A$ numbers)
lead to the unpredictable composition of the fission products,
which vary in a broad probabilistic and somewhat chaotic manner.
This distinguishes fission from purely quantum-tunnelling processes such as proton emission, alpha decay and cluster decay,
 which give the same products each time.

Second, while
 {\em $r$-process} production
  (from the lower to higher $A$ numbers)
  has been more studied and
 can be better modeled,
 the results strongly depend on the assumed equation of state (EOS) (\cite{f-a99}), 
 the neutron/seed ratio and the composition of the seed, which in models are characterized by the proton-to-nucleon ratio $Y_e$
 of the ejected and expanding matter.
 The value of $Y_e$ has basically two effects:
(1) It determines the neutron-to-seed ratio,
which finally determines the maximum nucleon number $A$ of the resulting abundance distribution, and
(2) it also determines the location (neutron separation energy) of the $r$-process path,
and thus the $\beta$-decay half-lives encountered; this influences the process speed and the nuclear energy release.
 Thus, $Y_e$ of the ejected matter strongly depends on how much crust or core matter is contained in the ejecta.
 Also, various processes such as neutrino transport, neutrino captures, or positron captures, alter $Y_e$ evolution.
 In neutron star merger studies, test calculations using different polytropic EOSs indicate a strong dependence of the amount of
ejecta on the adiabatic exponent of the EOS, where stiffer equations result in more ejected material. (\cite{f-a99})

Finally, the data on the abundance yields from the observed supernovae
are not useful for modeling the collision element production.
The two processes (supernova and collision) fundamentally differ in several aspects.
With respect to the nucleosynthesis reactions,   the two explosions
have
substantially different seed nuclei composition and neutron-seed ratios.
In supernova explosions, when the core collapses once coulomb repulsion can no longer resist gravity,
the propagating outward shockwave
causes  the temperature increase (resulting from the compression)
and produces a breakdown of
nuclei by photodisintegration, for example:
$^{56}Fe + \gamma \rightarrow 13 ^{4}He + 4 ^{1}n$, $^{4}He + \gamma \rightarrow 2 ^{1}H + 2 ^{1}n$.
The abundant neutrons produced by photodisintegration
are captured
by those nuclei from the outer layers (the ÓseedsÓ) that managed to survive.
Thus, the produced abundances depend strongly on the characteristics of the star.
Indeed,
astronomical observations confirm that supernova
nucleosynthesis yields vary with stellar mass, metallicity and explosion energy (see, for example, \cite{n06}).
 In comparison,
 the NRCO has no outer layers (other than $Fe$ crust)
 and thus does not follow the same, as in supernova, chain of photodisintegration reactions to supply the seeds and the free neutrons.
 On the other hand, in the collision, a completely different from supernova distribution of seeds is supplied by the nuclei from the target. 
But most importantly,
in the collision scenario, reactions of fission, rather then nucleosynthesis,
play the dominant role in the element production.

The only thing that can be said at this point is that,
in the framework of the outlined hypothesis,
the observed abundances of the solar system  represent the only outcome of such collision event known to us 
(of course, the final abundances also include contributions from 
stellar
and {\em in situ} sources). 
We do not have
a statistical sample to make any comparisons. 
If the fission and nucleosynthesis reactions were better understood,
the only subsequent approach would have been to solve the inverse problem, i.e.
to find out what the initial conditions had to be so the model resulted in the observed abundances.

\subsection{Probability of another collision event} \label{ss:probability}

We suspect that the collision event proposed in our scenario is a unique and extraordinary event.  
The only reason it was even envisioned  is because many existing characteristics of the solar system
are
otherwise
difficult to explain self-consistently (see Section~\ref{ss:puzzles}).

To get a feel for how likely  a similar event can be,
we estimate
the  time of a collision of an object with a target (star):
$\tau \sim L^3  /  (  \pi R^2 \times V$ ), i.e.,
space volume per star /  ( target cross-section $\times$  approach velocity). 
In the Sun's neighborhood, the average distance between the stars is roughly 5~light years ($L = 5 \times 10^{13} km$). 
Near the core of the Galaxy, 
stars are packed fairly tight, about 1/4 light year apart.
Because of gravity interaction, the paths of stars that would have missed each other
will curve toward each other and have a grazing collision even when their trajectories looked like they would miss initially,
thus increasing  the capture cross-section. 
So let us assume that the effective size of the collision target 
is of order of $R \sim 10^6 \, km$.  
Finally, assume the approach velocity $\sim 10^3 \,  km/s$. 
Thus, the characteristic time for one particular neutron star to collide with a star as
\begin{eqnarray}
\tau_1  \sim  \frac{(5 \times 10^{13} \, km)^3 }{ 3.14 \times (10^6 \, km )^2 \times 10^3 \, km /s } \nonumber 
\sim 10^{18} yrs. \nonumber 
\end{eqnarray}
Some astrophysicists estimate that there may be as many as $10^8$ neutron stars in our Galaxy.  \citep{k98} 
Furthermore, non-rotating and non-accreting neutron stars which have cooled down are virtually undetectable by conventional means.
Gravitational micro-lensing experiments have detected a population of such objects in the galactic halo. 
These stars have sufficiently high initial velocity.  \citep{k98}

Using the estimate of $10^8$ neutron stars,
we estimate
the probability of a collision of {\em some} 
neutron star with {\em some} planetary system in our galaxy
as
\begin{eqnarray}
1 / \tau = N \times 1 / \tau_1 \sim 10^8 stars \, / 10^{18} yr \,  \sim 10^{-10} yr^{-1}. \nonumber
\end{eqnarray}
Which means that this kind of a collision is indeed a very rare event, with the odds of being "once in a lifetime"
(the lifetime of the Universe so far is $\sim 1.34 \times 10^{10}$ years).%
\footnote{
To interpret what it means, imagine that the lifetime of the Universe is equivalent to one day of a person's life.
The person intends to start buying a lottery ticket every day going forward, and the odds are 1 in $X$.
The fact that such collision {\em already} occurred according to our hypothesis, is equivalent to winning the lottery on the first day.
}

This result also implies that the exoplanetary systems are most likely formed according to the "normal" processes of planet formation.
In the framework of our hypothesis, the "rocky" exoplanets most likely will {\em not} have the same composition as our terrestrial planets.
Data on the actual differences would provide valuable insights into the validity of our hypothesis.
Unfortunately, at this point, the measurement techniques employed to detect exoplanets
are not yet capable of determining what the exoplanets are actually composed of.%
\footnote{
ÓRockyÓ does not necessarily mean they are enriched with $r$- or $s$-processed elements.
}
%
\begin{eqnarray}
\ \nonumber
\end{eqnarray}

\section{Summary}  \label{S:conclusion}

We advanced a hypothesis that 
the early solar system (initially possessing no terrestrial planets)  
became enriched with the observed and extinct $r-$, $s$-, and $p$-process elements  
predominantly as a result of its collision with a neutron star-like object 
(a "neutron-rich compact object", or NRCO). 
These elements subsequently formed dust grains, accreted into larger and larger rocky bodies, and eventually formed the modern terrestrial planets.

To eject its neutron-rich core matter into the solar system and produce the chemical elements via complex chains of nuclear reactions, 
the NRCO had to explode, at least partially.  
We identified that the key mechanism leading to the explosion was 
density decompression and subsequent nuclear reactions 
in the back of the NRCO's core, 
which resulted from the straight-line deceleration of the NRCO 
as it penetrated a dense massive solar system body located within the zone of current terrestrial planets.  
(Such straight-line deceleration and subsequent decompression do not occur 
in the familiar stellar events, such as supernova or neutron star mergers, 
and therefore has never been examined before.) 
As established in nuclear physics,  
in the phase state of "nuclear fog" -- where nuclear liquid and gas co-exist -- 
density decrease by a factor of $10^2$ or more can indeed take place.  
From this perspective, in our model the "strength" of the NRCO's deceleration becomes then defined not in the kinetic sense, but 
in the thermodynamic sense.  
If the initial phase state of the neutron liquid is rather close to the boundary of the unstable two-phase zone, 
even a small deceleration magnitude in the kinetic sense, can still trigger sufficient density decompression. 
In the unstable zone, density perturbations develop exponentially fast. 
Furthermore, nuclear reactions occur with even faster time scales ($t \sim 10^{-22} \div 10^{-15}$~sec). 
Thus, even a short-lived localized fluctuation of density to below $\rho_{drip}$ 
can trigger the cascade of nuclear reactions.  
Recent findings from the high-density / high-energy nuclear physics 
established that nuclear reactions of multi-fragmentation of hyper-nuclei and fission of super-nuclei 
are capable of releasing sufficient amounts of energy, which is needed for the explosion and 
ejection of the NRCO's core matter against its gravitational pull.

At this point, we have not established many constraints on the nature of the target.  
Its only critical characteristic so far is that it must have been able to sufficiently decelerate the NRCO.   
From this perspective, the target could have been a "super-Jupiter" rotating on the first orbit 
(i.e., between the Sun and Jupiter, as no terrestrial planets existed at that time), 
or a close binary companion of the Sun (if such existed), 
or the edge of the Sun (if the collision could occur without destroying the Sun). 
If the target was the "super-Jupiter" or the Sun's companion, 
its current absence implies that it was either destroyed or ejected as a result of the collision.

With respect to the NRCO, 
our model requires {\em a priori} only that it contains dense neutron liquid in its core and has a solid crust (as do neutron stars), 
and has such other characteristics that in the collision its density 
(in a localized spot) 
decompresses to the level when nuclear reactions can cascade.  
For example, the NRCO has to travel fast enough, so the collision has a "head-on", straight-line type deceleration, 
to create decompression in the back. 
In our analysis, we considered  the acquisition of the target mass by the NRCO as the primary deceleration mechanism, 
which therefore required that the NRCO be gravitationally-powerful and sized in such a way relative to the target,  
so the target had enough mass to decelerate the NRCO. 
In general, other mechanisms of deceleration (interaction of magnetic fields, for example) may also play significant if not dominant role. 
Overall, however, it is the aggregate deceleration magnitude that is crucial, regardless of which mechanism is  dominant. 
The closer the NRCO's initial equation of state (EoS) is to the liquid/gas phase-transition boundary, 
the smaller the deceleration magnitude is required for the sufficient decompression and subsequent nuclear reaction cascade. 
(We demonstrated that NRCO's with such characteristics can theoretically exist, 
and even have smaller sizes than traditional neutron stars.  
Therefore, the role of NRCO can be played perhaps by a piece of a neutron star, torn apart and catapulted by a distant black hole.)

Once a portion of the NRCO's core matter is ejected into the solar system, 
two main element production mechanisms  take place:  
{\em multi-fragmentation/ fission}  (split of nuclei with high $A$ numbers into several with lower $A$ numbers), 
and {\em nucleosynthesis} (formation of elements by nucleon capture from lower to higher A numbers).
In the collision scenario,  
hyper-nuclei fragmentation/ fission are most likely  dominant. 
(In supernova, nucleosynthesis is the main mechanism.)

Unfortunately, a more detailed analysis is constrained 
by the limitations in the current understanding 
of (1) the EoS of high-density nuclear matter and (2) the nuclear reactions for hyper- and super-nuclei. 
While significant advances have been made by nuclear physicists in recent years, 
further developments of theory and additional experimental data are needed.  
Therefore, we cannot at this point estimate the energy or mass of the ejecta, nor the produced abundances of the elements, 
nor the implications for the orbital dynamics of the solar system objects, nor the fate of the colliding objects.
Nonetheless, the proposed hypothesis presents a new framework from which the entire evolution of the solar system can be re-evaluated.  

The collision hypothesis has the advantage over all other currently considered theories 
(among which the supernova enrichment theory being one of the most commonly accepted so far),  
because the collision scenario is  capable of self-consistently explaining {\em all} cosmo-chemical inconsistencies (see Section~\ref{ss:puzzles}) 
by invoking {\em only one} event. 
It is theoretically apparent  that an ejecta of super-hot neutron-rich hyper-nuclei into the surrounding medium from the target, 
can produce {\em every} chemical element ever detected in the solar system.  
The planetary structure of the solar system (with two classes of planets and atypically wide, circular and non-resonant orbits of giants) 
also fits better within the collision scenario.  
Finally, the collision hypothesis does not contradict any of the already developed theories, but rather integrates them all 
from a different perspective.  
For example, all of the stellar enrichment models are valid conceptually, 
but the relative contributions from various mechanisms perhaps is less than have been assumed so far.


\appendix

\section{Brief overview of models of equations of state for nuclear matter and neutron stars}  \label{A:NSmodels}

As examples, we list several models of equation of state for dense matter.

\paragraph{Particles with interaction}

Equations of state obtained from a Yukawa-type interaction in the Hartree--Fock approximation are (see
\citep{st83}, p.~217)
\begin{eqnarray}
\rho = n m + \frac{3}{10} (3 \pi^2)^{2/3} \frac{\hbar}{m c^2} n^{5/3} \pm \frac{\pi g^2}{\mu^2 c^2} n^2 ,
\nonumber\\
p = K n^{5/3} \pm \frac{\pi g^2}{\mu^2} n^2\label{HFa}
\end{eqnarray}
\emph{Here}, $\mu = \hbar / m_g c^2 $ is the inverse of the Compton wavelength of the $g$-field quanta,
$g$ is the "charge" of the Youkawa--like interaction. Condition $\mu \sim 1.46 \, fm$ ($1 \, fm =
10^{-15} \, m$) requires the field quanta ($\pi$-meson) to have a mass $\sim 140 \, Mev$ (i.e.
comparable to pion mass). The experimental data give $g^2 / \hbar c \sim 10$. For $W$-boson
(weak-interactions) $\mu_W \simeq 2.45 \times 10^{-18} \, m$.  For nucleon $\mu_p \simeq 1.32 \, fm$. 

For high densities, the equation of state is \emph{hardened} due to the dominance of the "repulsive
core" Yukawa potentials. As $n \rightarrow \infty$, the equation of state satisfies (ibid, p.~211)
$p \rightarrow \rho c^2 \label{ext} .$
The sound speed approaches in this case  $c_s = (\partial p / \partial \rho)^{1/2} \rightarrow c$, in
contrast to an ideal relativistic gas, for which
$p \rightarrow 3^{-1} \rho c^2$ with $c_s \rightarrow 3^{-1/2} c.$

The Bethe-Johnson equation of state (ibid, p.~221), as an example of a variational calculation, proposes
the equation of state
\begin{eqnarray}
p = n^2 \frac{\partial}{\partial n} ( \frac{\epsilon}{n} ) = 364 \, n^{a+1} \, Mev \, fm^{-3}
\nonumber\\
\quad \mbox{with} \quad c^2_s = \frac{n^a}{1.01 + 0.648 n^a} c^2,
\end{eqnarray}
where $a = 1.54, \; 0.1  \leq  n  \leq 3 \; fm^{-3}$ or $1.7 \times 10^{14} \leq \rho \leq  1.1 \times
10^{16} \, g \, cm^{-3}$. Obviously, a causality violation exists, $p > \rho c^2$, at high
densities.

Like all many-body calculations of the equation of state, the Bethe-Johnson calculation is by no means
the final word on the issue (ibid, p.~224).

\paragraph{Super-dense substances}

The inner core composition of a neutron star is not exactly known due to insufficient knowledge of the
physics of strong interactions in super-dense substances \citep{f09}. 
It is not unlikely that the core consists of a nucleon-hyperon substance, pion condensate, quark-gluon
plasma, or some other exotic states. According to some authors, if the properties of the neutron star
crust ($\rho < 0.5 \rho_0$) are relatively well described by non-ideal plasma models, then for $\rho > 5
\rho_0$ the description of the properties of a super-nuclear-density substance is severely hampered by
both the incompleteness of laboratory data and the absence of a complete theory of super-nuclear-density
substances.

\paragraph{Quark-gluon plasma}

From \citep{f09}: "Quark-gluon plasma [184 - 187] constitutes the super-hot and super-dense form of
nuclear matter with unbound quarks and gluons which are bound inside hadrons at lower energies (Figs 13a
and 13b). The existence of QGP follows from the property of asymptotic freedom of QCD [188 - 191], which
yields a value of $1 - 10 \, GeV / fm^3$ for the energy density of a corresponding transition (recall 
that $1 \, \Gamma ev$ is approximately a mass of a nucleon, $m_p = 931 \, Mev$, and $1 \, fm = 10^{-13}
cm$). Detailed numerical calculations give the critical conditions for the emergence of QGP: $T^{qg}_c
\simeq 0.15-0.2 \, \Gamma eV$, or $T^{qg}_c \simeq (1.8 - 2.4) \times 10^{12} \, K$ (Figs 13d and 13f).

The initiation of this plasma manifests itself in an increase in the number of degrees of freedom - from
three inherent in the pion gas at low temperatures, $T < T^{qg}_c$, to $40 - 50$ inherent in the QGP for
$T > (1-3) T^{qg}_c$. Since the energy density, the pressure, and the entropy are approximately
proportional to the excited degrees of freedom of the system, a sharp variation in these thermodynamic
parameters in a small vicinity of $T^{qg}_c$ accounts for the large energy difference between the
ordinary nuclear substance and the QGP. Like our customary "electromagnetic" plasma, QGP may be ideal
for $T \gg T^{qg}_c$, and nonideal for $T \simeq (1-3) T^{qg}_c$. The corresponding nonideality
parameter - the ratio between the inter-particle interaction energy and the kinetic energy $E_cin$ in
this case is given in the form $\Gamma \simeq g^2 / a T$, where numerical constants of order unit are
omitted, the inter-particle distance $a \sim T^{-1} \; (a \sim 0.5 \,fm, \; T = 200 eV)$, and the strong
interaction constant $g \sim 2$".

\section{VdW--like equation of state} \label{A:NRCO}

As mentioned in Section~\ref{ss:NS}, 
the involved in the proposed collision NRCO
must have been compact and gravitationally-powerful to contain in its interior the highly dense neutron ÓfluidÓ 
(which is necessary for the eventual formation of the intermediate and heavy elements). 
Various models for the equation of state have been proposed (for neutron stars), 
but their reliability decreases with growing density, because the 
precise relativistic many-body theory of strongly interacting particles 
is not fully developed. 

Thus, in order to study the collision process, 
for the model of the NRCO we use  
the Skyrme-like approach.  
This approach 
(common in nuclear physics) 
describes pair-interactions of particles 
whose mutual attraction is taken into consideration, as well as the 
existence of repulsing core within the particles. 
The approach also assumes the validity of the following invariances:  
translation, Galilean, rotational, isospin, parity, and time-reversal.  
Moreover, some terms are added that cannot be derived from the invariance argument, e.g., spin-orbit interaction, or the 
3-body-type medium effect.  

Essentially, 
the particle interaction is described by density.  
This framework is expected to 
represent 
the collective dynamics.  
As a result, 
the equation of state for the system 
may be similar to 
Van der Waals (VdW) model due to their observed similarity \citep{k06}.

\subsection{Theoretical concept} \label{tc}

For such a model, the free energy of the system is given by
\begin{eqnarray}
    F(\rho , T) =
    A + \frac{3}{8} \xi \bigg( - \frac{8}{3} T \ln \bigg[ T^{3/2} \frac{(3 - \rho)}{3 \rho} \bigg] - 3 \rho \bigg),
    \label{eos:00}
\end{eqnarray}
where $\xi$ and $A$ are constants. Knowing free energy, pressure $p (\rho , T) = \rho^2 \partial_{\rho} F (\rho , T)$ can be determined from the thermodynamical equality $d F = - s d T - p d (\rho^{-1})$:
\begin{eqnarray}
    p(\rho , T) = \frac{3}{8} \xi \bigg( \frac{8 \rho T}{3 - \rho} - 3 \rho^2 \bigg).
    \label{eos:0p}
\end{eqnarray}
This is the VdW \emph{interpolation} equation of the state, 
 where the dimensionless quantities $p, \rho$ and $T$ are pressure, density and temperature 
 normalized by the critical values $\rho_c , \, T_c$ and $p_c$, 
  whose existence follows from 
  the given form of the free energy. 
  The critical density is determined from the conditions $(\partial_{\rho} p)_c =0$ and $(\partial_{\rho \rho} p)_c =0$. The critical parameters are not independent and are related by the relation $p_c / \rho_c = (3/8) c^2 (T_c / m_N c^2) \equiv (3/8) c^2 \xi$, 
where $c$ is the light speed,  temperature $T_c$ is measured in energetic units,  $m_N$ represents the mass of a particle of the system. 
Dimensionless parameter $\xi = (T_c / m_N c^2) \ll 1$.

The meaning of the expressions (\ref{eos:00}) and (\ref{eos:0p}) is as follows: In classical gases under normal conditions, interaction between particles (molecules, atoms) is weak. As the interaction (pressure) increases, the properties of the system differ more and more from the properties of the ideal gas, and finally the gas enters its condensed state -- liquid. In the liquid state, interaction between particles is great, and properties of this interaction strongly depend on the specific type of the liquid. This is the reason why general formulae, describing quantitatively properties of liquids, do not exist (see \cite{ll69}--\cite{ll59}).

However, it is possible to find some {\em interpolation} formula which can qualitatively describe the transition between the gas and liquid (as done in the Van der Waals model). Such formula must produce correct results in two limit cases. For rarified gases, it should converge into formulae correct for ideal gases. But as the density increases, it should incorporate the fact that the compressibility of the matter is limited. Such formula then would qualitatively describe the gas behavior in the transition state.

Expression~(\ref{eos:00}) satisfies such conditions.  At low densities, $\rho \ll 3$, Eq.~(\ref{eos:00}) becomes the expression for the free energy of the ideal gas;  at large densities  $\rho \rightarrow 3$, it shows that unlimited gas compression is impossible (at $\rho > 3$, the argument in the logarithm is negative; 
it is also obvious that $\rho$ cannot be negative.)

Eq.~(\ref{eos:00}) represents only one of the numerous possible interpolation formulae satisfying the posed requirements. 
There are no physical reasons to prefer one such interpolation over the others. 
But the VdW form (Eq.~\ref{eos:00}) is one of the simplest and easiest to work with.

The nucleon-nucleon interaction is 
composed of two components: 
a very short-range repulsive part and a long-range attractive part. 
The nuclear interaction is similar to VdW- interaction  within a 
molecular medium. In a sense, the phase transition in nuclear matter resembles the liquid-gas phase transition in classical fluids. However, as compared to classical fluids the main difference comes from the gas composition: for nuclear matter, the gas phase is predicted to be composed not only of single nucleons, neutrons and protons, but also of complex particles and fragments depending on temperature conditions.

According to 
\cite{s04} and \cite{g84},  
the nuclear equation of state (EOS) can be presented as follows:
\begin{equation}
p = a \rho - b \rho^2 + c \rho^3 ,
\end{equation}
where $a = k_B T, \, b = k_B T_c / \rho_c$ and $c = 2 k_B T_c / 6 \rho_c^2$. The coefficients $b$ and $c$ depend directly on the value of the critical temperature $T_c$ and the critical density $\rho_c$.

\begin{figure}
\centering
\includegraphics[width=11cm]{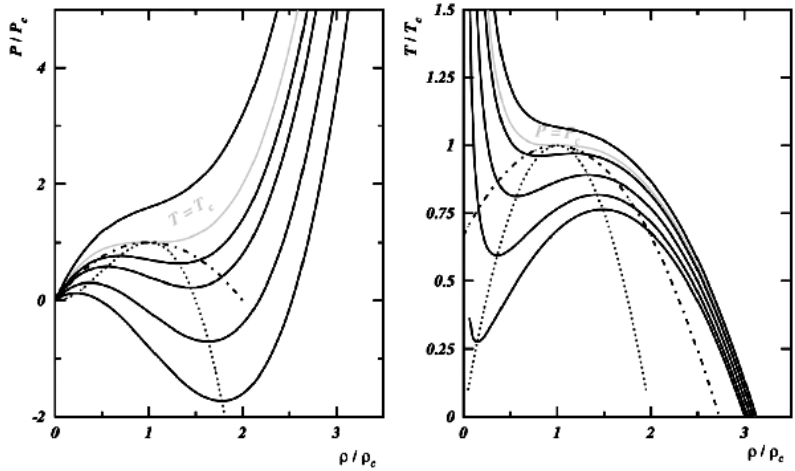}
\caption{
Equation of state for nuclear matter:  pressure (isotherms, left panel) or  temperature (isobars, right panel) as functions of  density.  
(Parameters are normalized to their critical values). 
The dash-dotted lines are the coexistence lines, the dotted lines are the spinodal lines.
From 
\cite{b02}.  
See also, \cite{br08}.}%
\label{br08}
\end{figure}

A typical set of isotherms for an equation of state (EoS) - pressure versus density with a constant temperature - corresponding to nuclear interaction  (Skyrme effective interaction and finite temperature Hartree-Fock theory, see 
\cite{j83}) 
is shown in Fig.~\ref{br08}. 

\begin{figure}
\centering
\includegraphics[width=8cm]{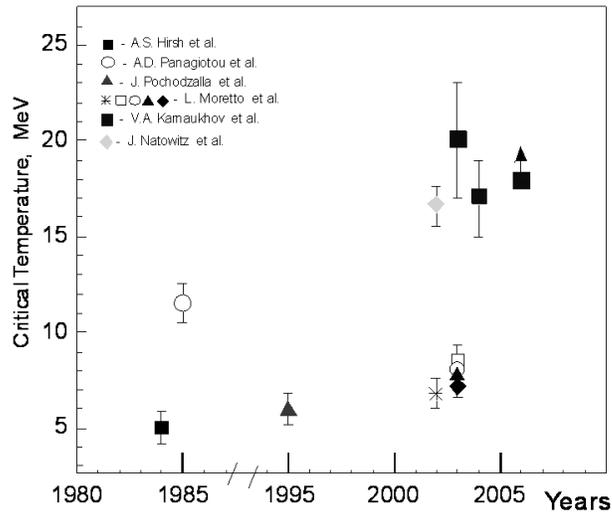} 
\caption{
Values of critical temperatures of nuclei $T_c$ measured by different techniques. From 
\cite{k-a11}.}%
 \label{TcKarn09}
\end{figure}

It exhibits the maximum-minimum structure typical of the VdW--like EoS. Depending on the effective interaction chosen and on the model 
(see \cite{j83}, \cite{j84}, \cite{c86}, \cite{m95}), 
the nuclear equation of state exhibits a critical point at $\rho_c \simeq (0.3 \div 0.4)\rho_0$ and $T_c \sim 5 \div 18 \, MeV$ 
(\cite{k09}, \cite{k-a11}). 
Calculations of $T_c$ were performed in 
\cite{s04}, \cite{g84}, \cite{s76}, \cite{j83}, \cite{z96}, \cite{t04}. 
 Experimental data are presented in Fig.~\ref{TcKarn09}.

In Fig.~\ref{br08}, $\rho_0$ refers to normal density. The region below the dotted line in Fig.~\ref{br08} corresponds to a domain of negative compressibility: here, at constant temperature, an increase of density is associated with a decrease of pressure. Therefore in this region a single homogeneous phase is unstable and the system falls into a liquid-gas phase in equilibrium. It is the so-called spinodal region, and spinodal instability produces  breaking into the two phases. Such instability has been proposed, for a long time, as a possible mechanism responsible for multi-fragmentation 
(\cite{b83}, \cite{h88}, \cite{l89}). 
The spinodal region constitutes the major part of the coexistence region (dashed--dotted line in Fig.~\ref{br08}) which also contains two meta-stable regions: one at density below $\rho_c$ for the nucleation of drops and one above $\rho_c$ for the nucleation of bubbles (nuclear cavitation). 
The boiling temperature for nuclear matter is $\sim 3$~times less then the critical temperature 
($T_b \sim 6 MeV$, $T_c \sim 18 MeV$, see \cite{k-a11}). 

The hot piece of nuclear matter produced in any nuclear collision has at most a few hundred nucleons $A \sim 200$ and so is not adequately described by the properties of infinite nuclear matter (or with $\ln \, A \gg 1$) where surface and Coulomb effects can not be ignored. These effects can  lead to a sizable reduction of the critical temperature.

Experimentally, it was shown that 
in a collision $p + Au$, first 
a combined excited 
nucleus is formed.  
The excitation energy 
is distributed between the thermal, compressional, and rotational (for heavy ions) energies. 
For colliding protons with initial $E =  8.1 \, MeV$, i.e., $T \sim 5 - 7 \, MeV$, the thermal energy component is $E_{th} > 400 \, MeV$. 
The nucleus "boils up" (see \cite{k-a11}; \cite{pk90}), 
and 
multi-fragmentation ($m$ fragments with $Z \sim (2 -20)$) and emission of protons and neutrons take place.

\subsection{Interpolating model of EoS}

The high density of the NRCO's nuclear matter requires that the VdW-like equation of state (EoS) be expanded into the relativistic domain. 
When $\rho \rightarrow 3$, the speed of sound, in the framework of model Eq.~(\ref{eos:0p}), tends to infinity, 
which is impossible because the sound speed is  always less then speed of light.  
For this reason,  
we require that 
the equation of state for the NRCO matter is analogous to the EoS of the VdW gas for small $\rho < 3$, but 
produces 
 in the relativistic density domain ($\rho \gg 3$) the sound speed equal to 
$1/\sqrt{3}$ 
(appropriate for the gas composed of free relativistic particles)\footnote{The current theory of quark-quark interactions suggests that quark interactions become arbitrary weak as the quarks are squeezed closer together (asymptotic freedom). In \cite{c-al74}, it was suggested that at very high densities, quark matter may be treated in the leading approximation as an ideal, relativistic Fermi gas, in which case  $p \rightarrow 3^{-1/2} c^2 \rho$ when $\rho \rightarrow \infty$. (\cite{st83}, p.~239,)}. 
Then, the interpolation formula for the free energy can be proposed in the form
\begin{eqnarray}
    F(\rho , T) = \nonumber\\
    \bigg( A + \frac{3}{8} \xi ( - \frac{8}{3} T \ln[ T^{3/2} \frac{(3 - \rho)}{3 \rho} ] -
       3 \rho ) \bigg) \theta [3 - x - \rho ] +
       \nonumber\\
       \frac{1}{3} \bigg( -  \ln  \frac{1}{\rho} - \frac{\alpha}{\rho} - \frac{\beta}{\rho^2} - \frac{\gamma}{\rho^3} \bigg) \theta [ \rho - 3 + x ]. \quad
    \label{eos:01}
\end{eqnarray}
Here, $\theta [s]$ is the Heaviside step--function, $\theta[s] = 1$ when $s > 0$ and $\theta[s] = 0$ when $s < 0$; 
free parameters $A, \alpha, \beta, \gamma, x$ 
assure the continuity of  free energy $F$; pressure $p = \rho^2 \partial_\rho F$; sound speed (squared) $s^2 = \partial_\rho p$. 
Parameter $x$ represents the transition point between relativistic and non-relativistic domains in the expression defining the free energy $F$.

In the domains of small and large densities, the speed of sound (squared) takes form: 
\begin{eqnarray}
     (s_-^2 )\equiv s^2 \bigg|_{\rho < 3 - x} = \frac{3 T \xi}{3 - \rho} - \frac{9 \xi \rho}{4} + \frac{3 T \xi \rho}{(3 - \rho)^2} \label{eos:02a},
     \\
     (s_+^2 )\equiv s^2 \bigg|_{\rho > 3 - x} = 6 \frac{\gamma}{\rho^4} + 4 \frac{\beta}{3 \rho^3}. \label{eos:02b}
\end{eqnarray}
Derivative $\partial (s_-^2 )$, obtained from Eq.~(\ref{eos:02a}), gives
\begin{eqnarray}
   \partial_\rho (s_-^2 ) = - \frac{9 \xi}{4} + \frac{6 T \xi}{(3 - \rho)^2} - \frac{6 T \xi \rho}{(3 - \rho)^3}
    \label{eos:03}
\end{eqnarray}

Eqs.(\ref{eos:01})-(\ref{eos:03})  form the system of equation which allows derivation of the EoS $p(\rho, t)$.
Because this system cannot be solved analytically, 
we solve it numerically and present results for one set of representative initial parameters.

For the following numerical example, we take values of  $\xi = 0.007$ and   $T = 0.1$,
which are within their commonly assumed ranges, 
but allow us to produce illustrating figures with good visual resolution. 

Next, we take small  $(s_-^2 )= 10^{-2}$ (i.e. the matter is near the phase transition boundary but is still in its liquid state) 
and replace $\rho \rightarrow 3 - x$.   
We find then that $x = 0.3489$ and $\partial_{\rho} (s_-^2 ) = 0.2809$.
Then
\begin{eqnarray}
  (s_+^2 ) = \frac{1}{3} - 0.0949 \beta - 0.1074 \gamma .
    \label{eos:04}
\end{eqnarray}
\begin{eqnarray}
\partial_{\rho} (s_+^2 )= 0.0716 \beta + 0.1215 \gamma.
 \label{eos:05}
\end{eqnarray}
Because the values of $s_{\pm}^2$ and of their derivatives are by definition the same at the transition point $x$, 
we can then solve for $\beta = 2.3745$ and $\gamma = 0.9140$. 
 Therefore, 
 the interpolation formula for the square of sound speed  $s_+^2$ becomes
 \begin{eqnarray}
s^2 = \bigg( \frac{0.0021}{(3 - \rho)} - 0.0158 \rho + (\frac{0.0021 \rho }{(3 - \rho)^2} \bigg)\theta [2.6511 - \rho]
    \nonumber\\
    + \bigg( \frac{1}{3} - \frac{1.8280}{\rho^3} - \frac{1.5830}{\rho^2} \bigg) \theta[- 2.6511 + \rho ]. \quad
 \label{eos:06}
\end{eqnarray}
Parameters $A$ and $\alpha$ are found from the continuity condition for the free energy and the pressure, 
resulting in 
$A = -0.1423 (-1.4921 + 0.8837 \alpha)$ and $\alpha =  - 4.9507$.
Thus, the final expression for the equation of state (EoS) becomes 
\begin{figure}
\centering
\includegraphics[width=8cm]{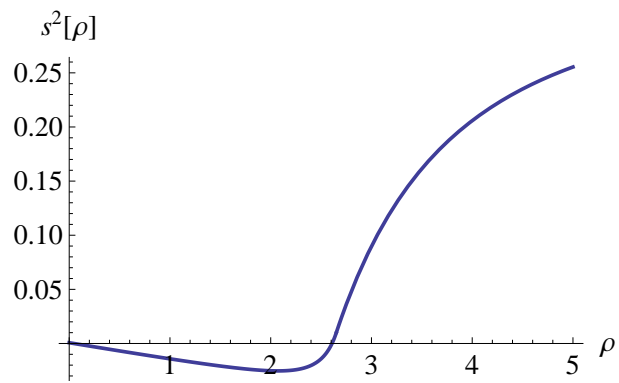} 
\caption{Dimensionless sound speed within the nuclear matter,  $s^2 (\rho)$, calculated for initial temperature $T = 0.1$.
When  $\rho \rightarrow \rho_0 \simeq 2.6$, the sound speed in the matter tends to zero. For $\rho < 2.6$, the system becomes unstable with respect to small fluctuation, and transits to two--phase state. \label{s2}}
\end{figure}
\begin{eqnarray}
p (\rho , T=0.1) =
\bigg( \frac{0.0021 \rho }{3 - \rho } - 0.0079 \rho^2  \bigg) \theta[2.6511 - \rho] + \nonumber\\
\bigg( \frac{\rho}{3} - 1.6502 + \frac{1.5830}{\rho} + \frac{0.9140}{\rho^2} \bigg) \theta[-2.6511 + \rho] \bigg). \quad
 \label{eos:07}
\end{eqnarray}
Obviously,  Eqs.~(\ref{eos:06}) and (\ref{eos:07}) can be generalized for 
arbitrary temperatures.

\begin{figure}
\centering
\includegraphics[width=8cm]{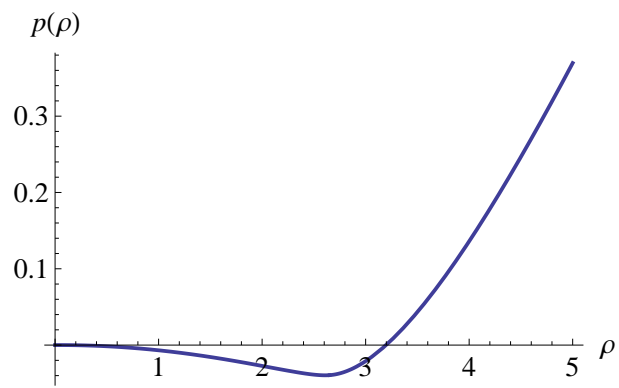} 
\caption{Dimensionless pressure.  \label{p}}
\end{figure}

\subsection{Radially-symmetrical distribution of mass}

To apply this framework to the spherical NRCO, 
first we demonstrate 
that the spherical configuration can indeed exist with the above-derived equation of state (VdW-like EoS with relativistic extension). 

We introduce quantity $m$ (here defined in usual units, but later normalized) which has the meaning 
of the "mass inside radius $r$":  
\begin{eqnarray}
m (r)  = 4 \pi \int_0^r d r r^2 \rho .
\label{eos:08}
\end{eqnarray}
The total NRCO mass, $M$, is the integral of (\ref{eos:08}) from $0$ to $R$, and 
includes all contributions to the mass including gravitational potential energy: in fact, the proper volume element in the gravity field is not  $4 \pi  r^2 d r$ but $4 \pi  r^2 g_{rr}^{1/2} d r = 4 \pi  r^2 (1 - 2 m / r)^{-1/2} d r$. 
When $r \rightarrow R$, $m$ must become equal to $M$, so that the interior metric 
matched smoothly the exterior Schwarzschild metric. (\cite{st83}) 

The equilibrium configuration is found from the Gilbert-Einstein's equations (Oppenheimer--Volkoff equations) written as
\begin{eqnarray}
\frac{d m}{d r} = 4 \pi r^2 \rho ,  \label{eos:09a}\\
\frac{d p}{d r} = - G \frac{m \rho}{r^2} ( 1 + \frac{p}{ c^2 \rho})(1 + 4 \pi \frac{p r^3}{m c^2}) ( 1 - 2 \frac{G m}{c^2 r} )^{-1},
\label{eos:09b}\\
\frac{d \Phi}{d r} = - \frac{1}{\rho} \frac{d p}{d r} ( 1 + \frac{p}{c^2 \rho} )^{-1} .
\label{eos:09c}
\end{eqnarray}
Terms $\sim c^{-2}$ and $\sim G / c^2$ describe  corrections produced by effects of the special and general theories of relativity.

Introducing 
dimensionless variables $r \rightarrow R_* r , \; \rho \rightarrow \rho_c \rho [r], \; m \rightarrow \rho_c R_*^3 m [r]$, 
we obtain that the (now dimensionless) quantity  $m [r]$  and its boundary condition become
\begin{eqnarray}
m [r] = 4 \pi \int_0^r d r r^2 \rho [r]
\label{eos:10a}
\end{eqnarray}
and
\begin{eqnarray}
m [1] = \mu \frac{M}{R^3} .
\label{eos:10b}
\end{eqnarray}
Here mass and radii are measured in the units of the Sun (denoted $\odot$), $M = M_* / M_\odot$, $R = R_* / R_\odot$, $M_\odot$ and $R_\odot$ are respectively the mass and radius of the Sun, dimensionless parameter $\mu = M_\odot / \rho_c R_\odot^3$. Numerically $\mu = 1.45 \times 10^{-13}$ if $\rho_c = 0.4 \times 10^{17} \, kg / m^3$.

Now we transform Eq.~(\ref{eos:09b}) to the form convenient for calculations
\begin{eqnarray}
\frac{d p [r]}{d r} = - \frac{G \rho_c^2 R_*^4}{p_c R_*^2} \frac{m [r] \rho [r]}{r^2} ( 1 + \frac{p_c}{c^2 \rho_c}
\frac{p [r]}{\rho [r]})\times \nonumber\\ (1 +
\frac{4 \pi p_c R_*^3}{\rho_c R_*^3 c^2}\frac{p [r] r^3}{m[r]}) ( 1 - \frac{2 G \rho_c R_*^3 }{c^2 R_*} \frac{m [r]}{r} )^{-1},
\label{eos:11a}
\end{eqnarray}
i.e.
\begin{eqnarray}
\frac{d p [r]}{d r} = - \chi R^2 ( \frac{c^2\rho_c}{p_c} ) \frac{m [r] \rho [r]}{r^2} ( 1 + ( \frac{p_c}{c^2 \rho_c} )
\frac{p [r]}{\rho [r]})\nonumber\\ \times  (1 +
4 \pi ( \frac{p_c}{\rho_c c^2} ) \frac{p [r] r^3}{m[r]}) ( 1 - 2 \chi  R^2 \frac{m [r]}{r} )^{-1}. \quad \quad
\label{eos:11b}
\end{eqnarray}
The universal 
coefficient $\chi = G \rho_c R_\odot^2 / c^2 = 1.4623 \times 10^9$. 
Parameter $\xi = T_c / m_N c^2$  varies from $0.005$ to $0.02$ depending on the assumed $T_c$. 
We use $\xi = 0.007$.

The resulting mass and pressure distributions are: 
\begin{eqnarray}
\frac{d m [r]}{d r} = 4 \pi r^2 \rho [r],  \label{eos:13a}\\
\frac{d p [r]}{d r} = -  \frac{8}{3 \xi}  \chi R^2 \frac{m [r] \rho [r]}{r^2} ( 1 + \frac{3}{8} \xi
\frac{p [r]}{\rho [r]})\nonumber\\ \times  (1 +
\frac{3 \pi}{2} \xi \frac{p [r] r^3}{m[r]}) ( 1 - 2 \chi  R^2 \frac{m [r]}{r} )^{-1}.
\label{eos:13b}
\end{eqnarray}
Together 
with Eqs.~(\ref{eos:06}) and (\ref{eos:07}) for the EoS  and the sound speed,   
they complete the system of equations from which 
a radially-symmetrical distribution of mass within the NRCO can be found. 
These equations 
cannot be resolved analytically in a general form. 
But their linearized version {\em can} be analytically solved (see next section). 

\subsection{Analytical solution}

To linearize Eqs.~(\ref{eos:13a}) and (\ref{eos:13b}), 
we write $\rho [r] = \rho [1] + \eta [r]$ and $m [r] = \frac{4 \pi}{3} \rho [1] r^3 + \mu [r]$. 
Quantities $\eta [r]$ and $\mu [r]$ may be considered as "add-ons", small perturbations of the basic state. 
The boundary conditions are $\eta[1] = 0$ and $\mu [0] = 0$. 
After substitution of these expressions into Eq.~(\ref{eos:13b}), 
we obtain the set of equation for the normalized density add-on $\eta [r]$ and mass add-on $\mu [r]$, in the linear approximation, 
\begin{eqnarray}
\frac{d \mu [r]}{d r} - 4 \pi r^2 \eta [r] = 0,  \label{eos:14a}\\
- \frac{2 \beta_{-6} \kappa R_{-6}^2 r \eta [r])}{( 1 - \beta_{-6}R_{-6}^2 r^2 )}  + \frac{d \eta [r]}{d r} =
- \frac{\alpha R_{-6}^2 r}{(1 - \beta_{-6}R_{-6}^2 r^2 )} \nonumber\\
- \frac{\lambda R_{-6}^4 }{(1 - \beta_{-6}R_{-6}^2 r^2 )^2} \mu [r]
- \frac{\nu R_{-6}^4 }{r^2(1 - \beta_{-6} R_{-6}^2  r^2 )} \mu [r].
\label{eos:14b}
\end{eqnarray}
Here, we used symbolism 
$R = 10^{-6} R_{-6}$ for notation brevity. 
For 
$T=0.1$, $\xi = 0.007$, $\chi = 1.4623 \times 10^9$,  and boundary conditions
\begin{eqnarray}
m [1] = 1.45 \times 10^{-13} \frac{M}{R^3}, \quad \rho [1] = 3.1970 \leftarrow p[1]= 0. \quad
\label{eos:13c}
\end{eqnarray}
numerical constants become 
$\alpha = 1.95 \times 10^{2}$, 
$\beta_{-6} = 3.92 \times 10^{-2}$, 
$\lambda = 0.57$, $\kappa = 7.78 \times 10^2$ and 
$\nu = 1.45 \times 10^1$. 
In Eq.~(\ref{eos:14b}), we neglected nonlinear terms of order $\eta^2$, $m^2$ and $\eta \, m$.
The square of sound speed in this state is $s^2 \simeq 0.12$.

The set of Eqs.~(\ref{eos:14a}) and (\ref{eos:14b}) for the normalized density add-on and mass add-on, 
produces the general solution which satisfies the boundary condition only if $R_{-6}^2 < \beta_{-6}^{-1} \simeq 25.5327$, i.e.,  radius $R_{-6}$ must satisfy condition  $0 < R_{-6} < 5.0530$. 
Thus, the normalized density add-on $\eta [r] $ can be expressed as 
\begin{eqnarray}
\eta [r] = \nonumber\\
\frac{\alpha R_{-6}^2 }{(1 - \beta_{-6} R_{-6}^2 r^2 )^{\kappa}} \int_r^1 d s  \, s (1 - \beta_{-6} R_{-6}^2 s^2 )^{\kappa - 1} +
\nonumber\\
\frac{\nu R_{-6}^2 }{(1 - \beta_{-6} R_{-6}^2 r^2 )^{\kappa}} \int_r^1 \frac{d s}{s^2} (1 - \beta_{-6} R_{-6}^2 s^2 )^{\kappa - 1} \mu [s] +
\nonumber\\
\frac{\lambda R_{-6}^4 }{(1 - \beta_{-6} R_{-6}^2 r^2 )^{\kappa}} \int_r^1 d s (1 - \beta_{-6} R_{-6}^2 s^2 )^{\kappa - 2} \mu [s].
\nonumber\\
\label{eos:15}
\quad
\end{eqnarray}
Quantity $\mu [r]$ satisfies the integral equation written here in the symbolic form
\begin{eqnarray}
\mu [r] = ( \frac{4 \pi}{3} \alpha R_{-6}^2 ) f[r] +
\nonumber\\
( \frac{4 \pi}{3} \nu R_{-6}^2 ) \widehat{G}_1 [r, s] \mu [s]
+ ( \frac{4 \pi}{3} \lambda R_{-6}^2 R_{-6}^2 ) \widehat{G}_2 [r, s] \mu [s],
\label{eos:16}
\end{eqnarray}
where
\begin{eqnarray}
f [r, R_{-6}] =
\nonumber\\
\int_0^r d s \, s^4 (1 - \beta_{-6} R_{-6}^2 s^2 )^{\kappa - 1} F_1^2 [ \frac{3}{2}, \kappa, \frac{5}{2}, \beta_{-6} R_{-6}^2 s^2 ]
\nonumber\\
 + r^3 F_1^2 [ \frac{3}{2}, \kappa, \frac{5}{2}, \beta_{-6} R_{-6}^2 r^2 ] \int_r^1  d s \, s {(1 - \beta_{-6} R_{-6}^2 s^2 )^{\kappa - 1}} ,
 \nonumber\\
\label{eos:17} \quad
\end{eqnarray}
and the Green's kernels $\widehat{G}_{1,2}$  and $\mu [r]$ are expressed via the hypergeometric functions $F_1^2 [a, b, c, z]$. (\cite{as64}) 

\begin{figure}
\centering
\includegraphics[width=8cm]{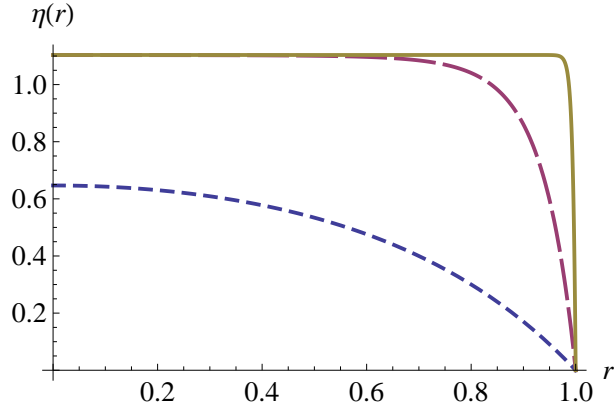} 
\caption{
Radial profile of density add-on $\eta (r) = \rho (r) - \rho [1]$ inside a NRCO with size:  
$R = 0.1$ 
(dash); 
$R = 0.3$ 
(long dash); 
$R = 1.0$ 
(solid).   
$R$ is measured in units $R_{-6} = 10^{-6} R_{\odot}$.
\label{densities}}
\end{figure}

Terms with $\nu $ and $\lambda R_{-6}^2 $ give small corrections to the leading (first) term in Eq.~(\ref{eos:15}) which is calculated analytically (can be also given by expressions):
\begin{eqnarray}
\eta_0 [r] =
(\frac{\alpha}{2 \kappa \beta_{-6}}) (1 - (\frac{1 - \beta_{-6} R_{-6}^2}{1 - \beta_{-6} R_{-6}^2  r^2 })^{\kappa})
\nonumber\\
\simeq (\frac{\alpha}{2 \kappa \beta_{-6}}) \bigg( 1 - \exp [ - \kappa \beta_{-6} R_{-6}^2 (1 - r^2 ) ] \bigg) <
\nonumber\\
(\frac{\alpha}{2 \kappa \beta_{-6}}) \bigg( 1 - \exp [ - \kappa \beta_{-6} R_{-6}^2 ] \bigg) (= \eta_0 [0]) <
\nonumber\\ (\frac{\alpha}{2 \kappa \beta_{-6}}) = 1.1036
\label{eos:18}
\end{eqnarray}
where $\kappa \simeq 2252 \gg 1 , \, \kappa \beta_{-6} \simeq 88.20 \gg 1 $ (Fig.~\ref{densities}).

\begin{figure}
\centering
\includegraphics[width=8cm]{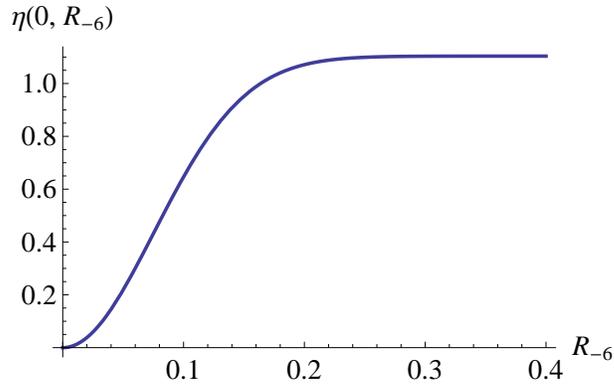} 
\caption{Density add-on $\eta (0, R_{-6})$ at the NRCO center ($r=0$) as a function of NRCO's size ($R_{-6}$). \label{denscenter}}
\end{figure}

The resulting density add-on at the center ($r=0$) of the NRCO, $\eta (0, R_{-6})$, 
depends on radius $R_{-6}$ as shown 
in Fig.~\ref{denscenter}.  
When NRCO's radius is small, not much density add-on occurs.  
The center becomes denser, as the radius of the NRCO increases  
up to about  $0.2 \times R_{-6}$.  
Above $\sim 0.2 \times R_{-6}$,  the density add-on (and therefore the total density) at the center remains constant.  

Therefore, the spherical configuration of the dense neutron matter composing the interior of the NRCO
can indeed exist with the VdW-like equation of state with relativistic extension 
(produced by the system of equations~(\ref{eos:01})-(\ref{eos:03})).

\subsection{Mass-Radius dependence for the NRCO}

The second boundary condition, Eq.~(\ref{eos:10b}), permits finding the mass-radius relationship. Thus, from
\begin{eqnarray}
m [r] = \frac{4 \pi}{3} \rho [1] r^3 + \mu [r]
\label{eos:19}
\end{eqnarray}
and boundary condition at $r=1$, it follows that 
\begin{eqnarray}
m [1] = \frac{4 \pi}{3} \rho [1]  + \mu [1] = \mu \frac{M}{(10^{-6} R_{-6})^3}.
\label{eos:20}
\end{eqnarray}
Here,
\begin{eqnarray}
\mu [1] = ( \frac{4 \pi}{3} \rho [1] \alpha  R_{-6}^2 ) \times
\nonumber\\
\int_0^1 d s \, s^4 (1 - \beta_{-6} R_{-6}^2 s^2 )^{\kappa - 1} F_1^2 [ \frac{3}{2},\kappa, \frac{5}{2}, \beta_{-6} R_{-6}^2 s^2 ].
\label{eos:21}
\end{eqnarray}
Solution of Eq.~(\ref{eos:20}) is 
\begin{eqnarray}
M ( R_{-6} ) \simeq 9.237 \times 10^{-5} R_{-6}^3  + 2.889 \times 10^{-5} \alpha R_{-6}^5 f[1, R_{-6} ] ).
\nonumber\\
\label{eos:22}
\end{eqnarray}
Here
\begin{eqnarray}
f[1, R_{-6}] =
\nonumber\\
\int_0^1 d s \, s^4 (1 - \beta_{-6} R_{-6}^2 s^2 )^{\kappa - 1} F_1^2 [ \frac{3}{2}, \kappa, \frac{5}{2}, \beta_{-6} R_{-6}^2 s^2 ].
\nonumber\\
\label{eos:23} \quad
\end{eqnarray}
This integral is calculated analytically and is expressed via a combination of hyper-geometrical functions $F_1^2 [a, b, c, z]$.
Numerical constants in Eqs.~(\ref{eos:22}) and (\ref{eos:23}) are approximately $\alpha = 1.95 \times 10^{2}$, $\beta_{-6} = 3.92 \times 10^{-2}$, $\kappa = 7.78 \times 10^2$ and $\rho [1] = 3.197 $.

\begin{figure}
\centering
\includegraphics[width=8cm]{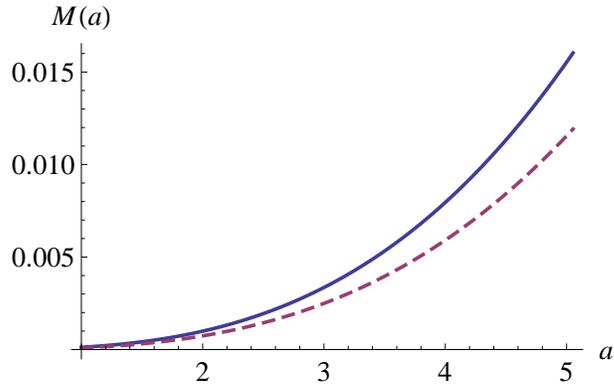} 
\caption{NRCO's mass $M$ as a function of its size $R_{-6}$   
for two cases: with relativistic correction (solid) and without (dash). 
\label{massaa}}
\end{figure}

Fig.~\ref{massaa} plots NRCO's mass as a function of radius, $M (R_{-6})$, for two cases -- with and without the relativistic correction.     
Notably, the model has a limit, 
it occurs approximately when   
$M \sim 0.016$, $R_{-6} \sim 5$.  
Beyond that, at least for this particular set of chosen parameters $\xi = (T_c / m_N c^2)  = 0.007$ and $T = 0.1$, 
the model no longer applies.  

\subsection{Summary of the results}

We developed a model for the VdW-ike equation of state  with a relativistic extension 
for a neutron-rich compact object (NRCO) 
composed of a highly dense neutron matter.

We presented a numerical solution  for one set of the determining parameters,  $\xi = (T_c / m_N c^2)  = 0.007$ and $T = 0.1$, 
which are within their commonly assumed ranges. 
The resulting model reveals the following characteristics of the NRCO:

 Within the spherical NRCO, the density profile exhibits density increase 
(characterized in our model by the density "add-on" $\eta$) toward the center of the NRCO.  

For NRCOs with small radii ($R \le 0.2 \times 10^{-6} R_{\odot}$, in usual units), 
the overall density profile inside a NRCO exhibits a smooth maximum at the center (Fig.~\ref{densities}). 
The density add-on $\eta$ at any particular distance from the center increases as the NRCO size $R$ increases (Fig.~\ref{denscenter}).  

For NRCOs with larger radii  ($0.2 \times 10^{-6} R_{\odot} < R \le 5 \times 10^{-6} R_{\odot}$, in usual units), 
the density distribution inside the NRCO is radially quasi-homogeneous, 
except for the narrow region near the edge where density gradient  $\rho ' ({\mathbf r}$) is negative and very steep. 
(See Fig.~\ref{densities} and \ref{denscenter}.) 

Notably this model allows for the existence of NRCOs with smaller sizes than the traditionally assumed sizes of neutron stars. 
Indeed, the model is valid (for this set of $\xi$ and $T$) for NRCOs with sizes up to 
$R \le 5 \times 10^{-6} R_{\odot}$ (in usual units), i.e., $R < 4 \, km$.  

\section{Decompression Model}  \label{A:decompression}

\begin{figure}
\centering
\includegraphics[width=6.5cm]{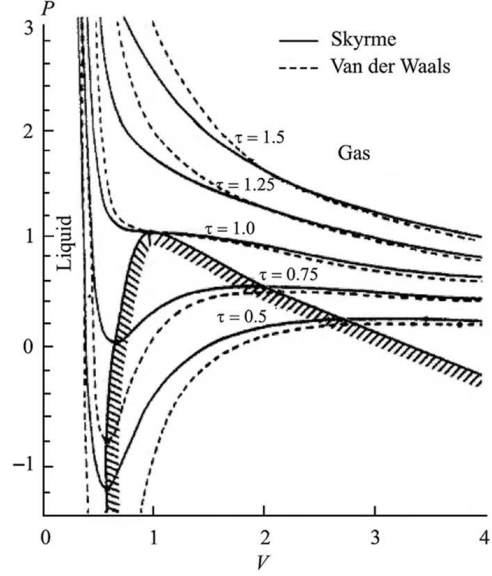}
\caption{
The equations of state $P(V)$ for a nuclear system interacting through a Skyrme potential and a Van der Waals compressible liquid-gas system  (shown in relative units). (From~\citet{j83}.)}
\label{nucl-phase2}
\end{figure}

\begin{figure}
\centering
\includegraphics[width=7cm]{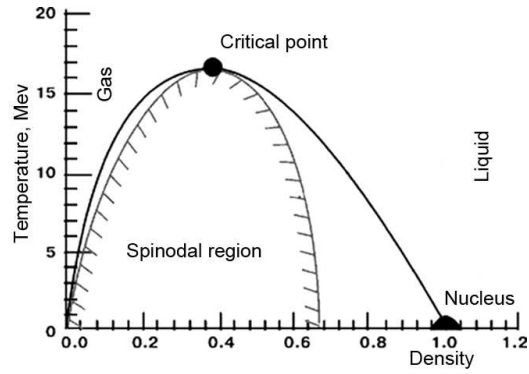} 
\caption{Theoretical $T(\rho)$ phase diagram for nuclear matter (adapted from \citep{k06}). The solid line is determined by condition $\partial p / \partial \rho = 0$ and marks the phase transition zone. Density is expressed in units of  $\rho_{nucleus} \simeq 2.85 \times 10^{14} \, g/cm^3 $. Temperature is expressed in $Mev$ units ($1 \, Mev \simeq 10^{10} K$).}
\label{spinodal-reg}
\end{figure}

The equations of state of a multi--body system of nucleons interacting via Skyrme potential 
(an analogy for the "giant nucleus" of a NRCO) 
is presented in Fig.~\ref{nucl-phase2}.  (From~\citet{j83}.) 
The very steep part of the isotherms (on the left side) corresponds to the liquid phase. The gas phase is presented by the right parts of the isotherms where pressure is changing smoothly with increasing volume. Of special interest is the part of the diagram where the isotherms correspond to the negative compressibility, i.e. $(\partial P / \partial V )_T > 0$. This is the so-called {\em spinodal} zone where the matter phase is unstable and can exist in both liquid and/or gas states.
Within the spinodal zone lies  a particularly unstable two-phased region (marked by the hatched line in Fig.~\ref{spinodal-reg}), in which random density fluctuations lead to almost instantaneous collapse of the initially uniform system into a mixture of two phases -- for nuclear matter, it is either liquid droplets surrounded by gas of neutrons, 
or homogeneous neutron liquid with neutron-
gas bubbles.

\begin{figure}[h!]
\centering
\includegraphics[width=7cm]{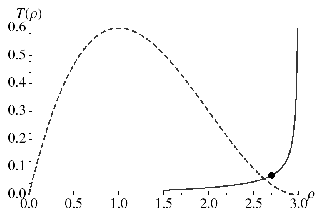} 
\caption{$T(\rho)$ phase diagram. Decompression path (solid line) of a localized pocket of the interior matter of the NRCO core. Depending on the initial state (solid dot) and the intensity of deceleration causing decompression, the state of the matter can shift into the two-phased region (area under dashed line) where the system cannot exist in the homogeneous state and forms nuclear fog -- a two-phase mixture of nuclear liquid and nuclear vapor. The boundary of the two-phased zone is defined by condition $\partial p / \partial \rho = 0$ (the zero speed of sound condition).}
\label{VDW-Trho}
\end{figure}

Critical temperature $T_c$ for the liquid-gas phase transition is a crucial characteristic 
of 
the nuclear equation of state. 
The nuclear equation of state (EoS), among other models, can be expressed  (see \cite{k09})  
as follows:
\begin{eqnarray}
p = a \rho - b \rho^2 + c \rho^3,
\label{kar01}
\end{eqnarray}
where $a = k_B T_c (T / T_c)$, $b = (k_B T_c) /\rho_c$ and $c = 2 (k_B T_c) / 6 \rho_c^2$. 
Coefficients $b$ and $c$ depend directly on the value of  critical temperature $T_c$ and  critical density $\rho_c$. 
This EoS is similar to Van der Waals equation suggested in 1875. 
For temperatures below the critical one, 
two distinct nuclear phases 
exist, and may 
coexist - liquid and gas. 
Above
$T_c$, 
the two-phase equilibrium  
does not exist, 
only the nuclear vapor does. 

It follows from Eq.~(\ref{kar01}) that the square of sound speed in the nuclear matter, 
$s^2 = \partial p / \partial \rho = a - 2 b \rho + 3 c \rho^2$, can be negative in some domain of densities, if temperature $T$ is small. 
However, the model~(\ref{kar01}) needs to be modified for very large densities to not violate the principle of causality ($s < c_{light}$).

There 
exist many numerical estimates of 
$T_c$ for finite nuclei. 
But results depend strongly on the chosen model. 
For example, models 
using the Skyrme effective interaction and the thermal Hartree-Fock theory, 
employing the semi-classical nuclear model, 
based on the Seyler-Blanchard interaction, 
and several others, have been considered.  
Experimental measurements 
of the critical temperature of the nuclear liquid-gas phase transition 
produced a range: 
$5 \, Mev < T_c < 18 \, Mev$ 
 (\cite{k09}).

The nuclear state where liquid and gas phases coexist, plays a crucial role in our collision scenario.  
As the NRCO decelerates and its interior matter pushes forward within the solid crust shell, the frontal part of the interior matter compresses, while the rear decompresses. The ($T, \rho$)-state of each individual localized "increment" in the rear part can be visualized as "sliding" down the decompression line on the $T(\rho)$-diagram (Fig.~\ref{VDW-Trho}, dashed line) from its initial state to its final state depending on the intensity of deceleration causing decompression. If the "sliding" path crosses the boundary of the two-phased region, the matter enters the zone of "nuclear fog" where the gas and liquid phases coexist. Once in that state, individual sub-regions within the "increments" may become significantly rarified to allow 
beta-decay and subsequent 
$r$-processes and fission. (See Section~\ref{ss:explosion}).
The key question is thus whether the collision deceleration can produce meaningful density decompression to shift the state of the matter into the unstable zone.

\subsection{Model of compression/decompression of the NRCO's interior \label{A:4}}

Equations of motion of classical fluid in the frame moving with acceleration ${\mathbf a} (t)$, are
\begin{eqnarray}
    \partial_t \rho + \partial_j \rho u_j = 0, \nonumber\\
    \rho \partial_t u_i + \rho u_j \partial_j u_i + \partial_i p = \rho g_i + \rho a_i (t),
     \nonumber\\
    \partial_j g_j = - 4 \pi G \rho .
    \label{f01}
\end{eqnarray}
Eqs.~(\ref{f01}) must be completed by boundary conditions.
Here $\rho$ is coordinate- and time-dependent density of the nuclear matter (not to be confused with
definitions from prior sections), $p = p ( \rho ) $ is pressure within the nuclear fluid, ${\mathbf u}$
is collective fluid velocity, ${\mathbf g}$  is gravity acceleration generated by the distributed mass,
${\mathbf a} (t)$ is the externally-imposed deceleration (which does not depend on coordinates):
$a_i (t) = - ((c_1)_c / \tau) \mu ' [((c_1)_c \tau /L_n ) t]$. In the last expression, in contrast to Eqs.~(\ref{f01}),
$t$ is the dimensionless time normalized on some $\tau$, prime signifies the derivative with respect to argument. For
simplicity, the evolution process is assumed to be adiabatic (entropy per unit mass is constant).

By multiplying the first Eq.~(\ref{f01}) on $ v_i $ and by combining the obtained equations,
we obtain
\begin{eqnarray}
    - \partial_t^2 \rho + \Delta (p + \rho v_i v_j ) = \partial_i ( \rho g_i ) + a_i (t) \partial_i \rho ,
 \nonumber\\
 \quad \quad
    \partial_j g_j = - 4 \pi G \rho
    \label{f03}
\end{eqnarray}
In the equilibrium state, when the acceleration ${\mathbf a} (t) = 0$ and the hydrodynamical velocity
${\mathbf v} = 0$,
\begin{eqnarray}
    \Delta p_s  = \partial_i ( \rho g^{(s)}_i ) ,
 \quad
    \partial_j g^{(s)}_j = - 4 \pi G \rho_s
    \label{f05}
\end{eqnarray}
Letting $\rho_1 = \rho - \rho_s , \, p_1 = p - p_s$, we obtain the nonlinear equations
\begin{eqnarray}
    - \partial_t^2 \rho_1 + \Delta (p_1 + \rho v_i v_j ) =
    \partial_i ( \rho_s g^{(1)}_i ) +  \nonumber \\
    \partial_i ( \rho_1 g^{(s)}_i ) +
    \partial_i ( \rho_1 g^{(1)}_i )+ a_i (t) \partial_i \rho_s + a_i (t) \partial_i \rho_1
    \simeq \nonumber \\
    \partial_i ( \rho_s g^{(1)}_i ) +  \partial_i ( \rho_1 g^{(s)}_i ) + a_i (t) \partial_i \rho_s  ,
 \nonumber\\
 \quad \quad
    \partial_j g^{(1)}_j = - 4 \pi G \rho_1
    \label{f06}
\end{eqnarray}
By neglecting the secondary effects connected with self-gravity interaction and a hydrodynamical motion,
$v^2 \rightarrow 0$, we find the correlation between the leading terms in the equation
\begin{eqnarray}
    - \partial_t^2 \rho_1 + \Delta_1 p_1 \simeq a_i (t) \partial_i \rho_s  .
    \label{f07}
\end{eqnarray}
Switching to dimensionless variables $p \rightarrow p_c p$, $\rho \rightarrow \rho_c \rho $, $x_k
\rightarrow R_* x_k$, we obtain
\begin{eqnarray*}
    - \frac{\rho_c (c_1)^2_c }{L_n^2} \partial_t^2 \rho_1 + \frac{p_c}{R^2_*} \Delta_1 p_1 \simeq
    \rho_c \frac{1}{R_*} [ \partial_x \rho_s ] \bigg( - \frac{(c_1)^2_c }{ L_n } \mu ' (t) \bigg)  .
    \label{f09}
\end{eqnarray*}
Prime signifies the derivative with respect to argument. Thus,
\begin{eqnarray*}
    \epsilon \beta \partial_t^2 \rho_1
    + \Delta_1 p_1 \simeq
    \beta \, \rho_s ' \, \mu ' (t)  \, \frac{x}{\sqrt{x^2 + {\mathbf x}_\perp^2}} .
    \label{f10}
\end{eqnarray*}
Here, we introduced the small dimensionless parameters $\epsilon = R_* / L_n \ll 1$ and $\beta =  ({R_*
}/{L_n}) ({(c_1)_c }/{c })^2 ({\rho_c c^2 }/{p_c } )$. In leading approximation with respect to
$\epsilon$, Eq.~(\ref{f10}) becomes
\begin{eqnarray}
    ( \Delta_\perp + \partial_x^2 ) p_1 \simeq \beta \mu ' (t) [\rho_s ] '
    \frac{x}{\sqrt{x^2 + {\mathbf x}_\perp^2}} .
    \label{f11}
\end{eqnarray}
The solution of Eq.~(\ref{f11}) is expressed via the Green function of the problem $( \Delta_\perp +
\partial_x^2 ) G ( | {\mathbf x} - {\mathbf z} | )= \delta^{(3)} ({\mathbf x} - {\mathbf z})$ with zero
boundary condition for pressure $p_1 =0$, and the for  boundary condition $G ( | {\mathbf x} - {\mathbf
s} | ) = 0$ for the Green function on the surface of container $|{\mathbf s}| = 1$%
\footnote{
Consider $V$ as the internal region of a sphere $| {\mathbf s} | < R$. The Green's function satisfies the
equation
$$
\Delta  G ({\mathbf s}, {\mathbf r} ) = \delta ({\mathbf s} - {\mathbf r} ).
$$
Consider now the condition of $G ({\mathbf s}, {\mathbf r}) |_{|{\mathbf s}| = R} = 0$ on the surface of the sphere. The
Green's function of the problem is
$$
G ({\mathbf s}, {\mathbf r} ) = - \frac{1}{4 \pi} \bigg( \frac{1}{|{\mathbf s} - {\mathbf r}|}  -
\frac{R}{s} \frac{1}{|(R / s)^2 {\mathbf s} - {\mathbf r}|} \bigg).
$$}%
:
\begin{eqnarray}
    p_1 \simeq \beta \mu ' (t) \int d {\mathbf r}_{1 \perp} d x_1 \, G ( | {\mathbf x} - {\mathbf x}_1 | )
    \rho_s ' (|{\mathbf x}_1|)  \frac{x_1}{\sqrt{x_1^2 + {\mathbf r}_{1 \perp}^2}} . \;
    \nonumber \\
    \label{f12}
\end{eqnarray}
Eq.~(\ref{f12}) shows that the magnitude of $p_1$ is not identically zero. It can be positive or negative,
and depends strongly on the behavior of $\rho_s$ inside of the NRCO, especially near the gas-liquid phase transition boundary. 
Since $p_1 \sim \mu'$ (Eq.~\ref{f12}) and
 $p_1  = p (\rho_s + \rho_1) - p (\rho_s) \simeq s^2 (\rho_s) \rho_1$, 
it means that 
when $s^2 \rightarrow
0,$ the magnitude of density perturbation $\rho_1$ can reach 
very large (positive or negative) values.   

This means that when the NRCO deceleration $\mu '$ is sufficiently strong, 
significant decompression may indeed occur.

\section{Model describing deceleration of a NRCO due to accretion of surrounding nebula particles}
\label{A:accretion}

As mentioned, when a colliding object, like the NRCO, penetrates a target, multiple effects contribute to
dissipation of its kinetic energy  and cause deceleration. Depending on the properties of the target,
such effects as hydrodynamical drag, Cherenkov--like radiation of various waves related to collective
motions generated within the target, gravitational accretion of target particles onto the NRCO,
or distortion of the magnetic fields, may actually play meaningful roles. Here, we focus {\em only} on
the effect of accretion of the gaseous target particles onto the gravitationally powerful NRCO.%
\footnote{
Commonly, though, the accretion $\sim \dot{M}_*$ term in Eq.~(\ref{m1b}) is omitted from
consideration from the very start. This is often correct. One such case is, obviously, when the parts of
the system do not travel relative to each other and the star does not rotate, particle accretion onto
the star is spherically symmetrical. In this case,  averaged $\overline{{\mathbf v}} = 0$ and the
accretion term in Eq.~(\ref{m1b}) is zero. In general, however, when the star is moving relative to the
nebula, velocity $\overline{{\mathbf v}} $ is not necessarily zero and requires explicit calculations.
Without analyzing the specifics of the problem, it is not immediately obvious that the second term  can
be dismissed a~priori.
}

Accretion onto a supersonically moving star has 
been studied both theoretically and numerically.
(See review by \citet{ec04}.)
The early works of \cite{hl39} and  \cite{b52}  
considered accretion by a star moving at a steady speed through an infinite gas nebula.
Later, the problem was applied to accretion of particles from interstellar medium,
a stellar wind, or a common envelope
(where two stellar cores become embedded in a large gas envelope formed
when one member of the binary system swells)
\citep{p78, ts00, d64, b01, ec04, r94, bkp97, p00, t-a12}.
Accretion onto a neutron star from the supernova ejecta has also been extensively researched --
for a radially-outflowing ejecta
\citep{c71, zin72},
for an in-falling ejecta
\citep{c89, csw96}
and even when a star is moving at a high speed across the supernova ejecta
\citep{z-a07}.

The equation of motion for a body of variable mass follows from the law of conservation of linear
momentum of the {\it entire} system  composed of the object and the surrounding mass captured by the
object \citep{m97}\footnote{
The elementary demonstration of the basic equation follows from a calculation of difference of linear moments in final
and initial states. In modern notation, we obtain
$\Delta ( M_* {\mathbf V} ) = (M_* + \Delta M_*) ({\mathbf V} + \Delta {\mathbf V}) +
(- \Delta M_*) ({\mathbf V} + \Delta {\mathbf V} + {\mathbf V}_{rel}) - M_* {\mathbf V} = {\mathbf F} \,
\Delta t ,$
where $M = M_* + \Delta M_* + (- \Delta M_*)$, ${\mathbf V}_{rel}$ is a relative velocity of mass
$(-\Delta M_*)$ with respect to $M_* (t).$
}. Thus, when a neutron star enters a dense gaseous "cloud", and surrounding nebula
particles accrete onto the gravitationally powerful star, the motion of the star will be described by
$ d ( M_* {\mathbf V} ) - {\mathbf v} \, d M_* = \delta {\mathbf I}. $
Here $M_* (t)$ and  $\mathbf V(t)$  denote, respectively, the mass and  velocity of the moving star in a
inertial frame at instance $t$, ${\mathbf v(t)}$ is the velocity in the same frame of the accreting
nebula particles which constituent the composing mass $d M_*$), and $d (...)$ denotes change of
quantities over the interval of time $d t$. Quantity $\delta {\mathbf I} = M_* {\mathbf a} \, d t$
(where ${\mathbf a(t)}$ is acceleration) is the impulse of an external force ${\mathbf F} = M_* \,
{\mathbf a}$.
Then it follows:
\begin{equation}\label{m1b}
d {\mathbf V} + ({\mathbf V} - {\mathbf v}) \, \frac{ d M_*}{M_*}  = {\mathbf a} dt.
\end{equation}
Expression~(\ref{m1b}) must be statistically averaged with respect to all possible values of velocities
of the accreting particles  for the given $d M_*$. After the averaging, the velocity  ${\mathbf v}$ of
accreting fragment $d M_*$ in Eq.~(\ref{m1b}) which contains a large number of accreting particles, is
replaced by averaged $\overline{{\mathbf v}}$.

To derive the expression for  $\overline{{\mathbf v}} $, we assume that the surrounding gas is composed
purely of ionized hydrogen--degenerated electron--proton plasma.
Since $m_p \gg m_e$, only the proton component is significant for the star mass change.%
\footnote{
We assume that NS luminosity $L$ does not surpass the Eddington's limit, $L > L_e$ where $L_e$ is
obtained by setting the outward continuum radiation pressure equal to the inward gravitational force.
Both forces decrease according to the inverse square laws. Therefore, once equality is reached, the
hydrodynamic flow is constant throughout the star. The outward force of radiation pressure is given by
$\partial_r p = - c^{-1} \kappa \rho F_r = - c^{-1} \sigma_T \rho m_p^{-1} (L / 4 \pi r^2 ).$ Here,
$\kappa$ is the opacity of the stellar material, $\sigma_T$ is the Thomson cross--section for the
electron, and the gas is assumed to be composed purely of ionized hydrogen. The pressure support of the
star is given by the equation of hydrostatic equilibrium $\partial_r p = - \rho g = - G \rho M / r^2 .$
(See \citep{zn96}.)
}  %
%
 We assume spherical symmetry of the velocity distribution of the gas particles, so that distribution
function $f ({\mathbf v}) = f (v)$ is a function of velocity module. The probability that any gas
particle occupies element $d w = dv_x dv_y d v_z$ in the space of velocities is proportional to $d w \,
f (v).$ The probability of the star to capture the gas particle with velocity ${\mathbf v}$ is
proportional to the cross--section of interaction, i.e. to the product of the module of relative
velocity of the particle with respect to the star ($|{\mathbf v} - {\mathbf V}|$) and $d w \, f ( v )$.
Thus, the average velocity is
\begin{equation}\label{vm1}
\overline{{\mathbf v}} = \frac{\int d w \, {\mathbf v} \, |{\mathbf v} - {\mathbf V}| \, f (v)}{\int d w
\, |{\mathbf v} - {\mathbf V}| \, f (v)}
\end{equation}
Due to the axial symmetry of the problem, $\overline{{\mathbf v}} $ is co-linear with ${\mathbf V}.$ In
the spherical coordinate system with $d w = $ $2 \pi d \theta \, \sin \theta \, d v \, v^2$ where
$\theta$ is the angle between ${\mathbf v}$ and ${\mathbf V},$
\begin{eqnarray*}
\overline{v} = \nonumber \\
\frac{ 2 \pi \int_0^{\infty} \int_0^{\pi} d v \, v^2 \, d \theta \, \sin \theta \, f (v) v \, \cos
\theta \, \sqrt{v^2 + V^2 - 2 v V \cos \theta} }{ 2 \pi \int_0^{\infty} \int_0^{\pi} d v \, v^2 \, d
\theta \, \sin \theta \, f (v) \, \sqrt{v^2 + V^2 - 2 v V \cos \theta} } ,
\end{eqnarray*}
which, after integrating with respect to angle $\theta, $ produces the following expression:
\begin{eqnarray*}
\overline{v} = \nonumber \\
\frac{\int_0^V d v \, v^3 f (v) [\frac{2}{3} v - \frac{2}{15}
\frac{v^3}{V^2}] + \int_V^{\infty} d v \, v^3 f (v) [\frac{2}{3} V - \frac{2}{15}
\frac{V^3}{v^2}]}{\int_0^V d v \, v^2 f (v) [2 V + \frac{2}{3} \frac{v^2}{V}] + \int_V^{\infty} d v \,
v^2 f (v) [2 v + \frac{2}{3} \frac{V^3}{v^2}]} .
\end{eqnarray*}
For the distribution with respect to velocities of fully degenerated non--relativistic Fermi gas
\citep{f29, f62}, $f (v) \sim H ( v_F - v)$, where $v_F$ is the local Fermi boundary velocity of the
nebula particles and $H [\xi]$ is the Heaviside step function.

When temperatures of proton-- and electron--components of the nebula are of the
same order, parameter $v_F = (6 \pi^2 / 2)^{1/3} \hbar (\rho / m_p)^{1/3} / m_p$ is of the same order as
the speed of sound $c_n$ in the nebula, and we  can write it as $v_F = \sqrt{3}\,  c \, c_1 \rho^{1/3} $
where numerical coefficient $c_1$ combines most of the constants, and $c$ is the speed of light. This
allows us to simplify our subsequent analysis by introducing dimensionless velocity $\mu = V / v_F$ and
expressing $\overline{v}$ as
\begin{eqnarray}
\overline{v} = 
-v_F \frac{[-24 \mu^7 + \alpha_1 (\mu) H[-1+\mu] + \alpha_2 (\mu) H[1-\mu]}{\beta_1 (\mu)  H[-1+\mu] +
\beta_2 (\mu)[1-\mu])}
\nonumber \\
\equiv  v_F \Phi (\mu) . \quad \label{v-m2}
\end{eqnarray}
Here $\alpha_1 (\mu) = 4( 1- 7 \mu^2 + 6 \mu^7 )$, $\alpha_2 (\mu) = 7 \mu^3 ( -5 + 2 \mu^2 + 3 \mu^4
)$, $\beta_1 (\mu) = 7 \mu (-24 \mu^5 + 4(-1- 5 \mu^2 + 6 \mu^5)$ and $\beta_2 (\mu) = 5 \mu (-3 - 2
\mu^2 + 5 \mu^4)$ (For small $\mu$, $\Phi (\mu) \simeq - \frac{1}{3} \mu + \frac{16}{45} \mu^3$; for
large $\mu$, $\Phi (\mu) \simeq - \frac{1}{5 \mu} + \frac{12}{175 \mu^3}$).
Then Eq.~(\ref{m1b}) (without the term $\mathbf a$) takes form suitable for our analysis:
\begin{equation}\label{eqm1b}
\frac{d {\mathbf \mu}}{dt} =( - \frac{\dot{M}_*}{M_*} )(1 - \frac{1}{\mu} \Phi (\mu))
 {\mathbf \mu}
\end{equation}
In a simplified interpolating form, this expression for  $\overline{v} $ can be parameterized as 
\begin{equation}\label{b2c2}
\overline{v}  = v_F( -\frac{86.25 \mu}{3 \times 86.25 + 5 \times 96.59 \mu^2}) \, ,
\end{equation}
which is valid for cases of both small and large of the parameter $\mu = v / v_F$.

Next we express star mass $M_*(t)$ in terms of (constant) $M_*$ (which from this point on will be
labeling the \emph{initial} mass of the star) and the normalized variable  $m(t)$, so that  $M_*(t) =
M_* m(t)$. Nebula density will be denoted $\rho({\mathbf r}) = \rho_c \, \rho (t)$ where ${\mathbf r}$
is the radius vector from the center of the nebula, $\rho_c$ is the nebula density in its center. We
introduce $(c_1)_c$, sound speed in the nebula center. Note that $v_F = \sqrt{3} (c_1)_c \rho^{1/3}
(t)$. We also will transition from old $\mu = V/ v_F$ to new $\mu = V/(c_1)_c$ (which leads to
recalculation of coefficients in front of $\mu$). After all these manipulations, we then obtain the
dimensionless equation of motion for the compact star:
\begin{eqnarray}
\label{em1} \dot{\mu} (t) +
\nonumber\\
\bigg(1+ \frac{86.25 \rho^{2/3} (t)}{160.99 \mu^2 (t) + 258.75 \rho^{2/3}
(t) }\bigg) \mu (t) \frac{\dot{m} (t) }{m(t)}
= 0 .
\end{eqnarray}
Here $\tau$ is the time--scale and all time derivatives are dimensionless. Coefficient $(c_1)_c = c
(1.83 / c \sqrt{3})\times 10^3 (\rho_c /  \rho_{\odot} ) = 3.524 \times 10^{-6} \rho^{1/3} \, c$,
where $c$ is the speed of light, (constant) $\rho_c$ is the density of the target nebula at its center,
and we again will express nebula density in terms of a constant, $\rho_c$, and a new normalized variable
$\rho(t)$, so that $\rho = \rho_c \rho(t)$ or $(\rho_c / \rho_{\odot} ) \rho_{\odot} \rho(t)$, if
expressed in the units of the average density of the Sun  $ \rho_{\odot} = 1.384 \times 10^3 \,
kg/m^3$. Also, $R = R_* / R_{\odot}$ and $M = M_* / M_{\odot}$. Going forward, we will label $\rho
\equiv \rho_c / \rho_{\odot}$ and $\rho(t)$ will be the normalized dimensionless variable.

Eq.~(\ref{em1}) describing the evolution of $\mu (t)$ contains two more time-dependent variables, $m(t)$
and $\rho(t)$. Thus we need to develop evolution equations for each of them to complete our system of
equations.

As mentioned, nebula density is a function of coordinates. The most general mono--parametric expression
for it is $\rho ({\mathbf r}) = \rho_c F ({\mathbf r} /L_n)$.
It implies that maximum of density $\rho_c$ is located at the center ($r=0$) and density decays away
from the center over the characteristic scale $L_n$. Nebula density at the point of the compact star
location is then expressed as $\rho ({\mathbf r} (t) = \rho_c F ({\mathbf r}(t)/L_n)$, where ${\mathbf
r} (t)$ is radius vector to location of the star, $F (\xi)$ is some function of radius. To
dimensionalize the equation below, we write ${\mathbf r} (t) = L_n {\mathbf s} (t)$. Because derivative
$d {\mathbf r} / dt = {\mathbf V} (t)$, then $(L_n / (c_1)_c \tau ) \, d {\mathbf s} / dt = {\mathbf
V}(t) / (c_1)_c \equiv {\mathbf \mu}$. If time-scale $\tau$ is chosen as $\tau = L_n / (c_1)_c$, the
second equation of the system connecting variables $\mu(t)$ and $s(t)$ -- the implicit form of the
nebula density -- is then
\begin{eqnarray}
\label{em3} \dot{s} (t) - \mu (t) = 0. \quad \quad
\end{eqnarray}
To derive expression for the evolution of star mass $m(t)$, we employ expression for $ \dot{M}_*$ due to
accretion proposed by Bondi  \citep{b52}:
\begin{eqnarray}
\dot{M}_* \equiv \pi R^2_{ac} \, \rho_\infty \, V = 4 \pi \alpha (G M_* )^2 {\rho_\infty } {(V^2 +
c_\infty^2)^{- 3/2}} \label{b0}
\end{eqnarray}
(see also, \cite{st83}, p.~420).  Here, $\rho_\infty$ is the medium density at large distance from
accreting body, $V$ is the velocity of the accreting body, $c_\infty$ is the sound speed in the nebula
at large distance from the accreting body,
coefficient $\alpha$ is of the order of unity.

In terms of dimensionless quantities $m (t), \, \rho (t)$ with time scale $\tau = L_n / (c_1)_c )$, we obtain
for the nebula composed from the degenerated Fermi--gas
\begin{eqnarray}\label{b2b}
\frac{d m (t)}{dt}  =  ( 4 \pi \alpha  G^2  \rho_{\odot}  \frac{ M_{\odot} R_{\odot} }{
(c_1)_c^4} ) M L \rho
\times \nonumber \\
\frac{ m^2 (t) \rho (t) }{ ( \mu^2 (t) + \rho^{2 / 3} (t) )^{3 / 2}}.
\end{eqnarray}
Here, $M = M_* / M_{\odot}$ is the initial compact star mass measured in units of the Sun's mass
($M_{\odot} = 2 \times 10^{30} \, kg$), $L = L_n / R_{\odot} $ is the characteristic nebula size
(in units of the Sun's radius, $R_{\odot} = 7 \times 10^8 \, m$), $\rho = \rho_c / \rho_{\odot}$
is the density of the nebula in its center (in units of the average density of the Sun,
$\rho_{\odot}= 1.38 \, g/cm^3$). Dimensionless accretion parameter
$k_A = 4 \pi \alpha  G^2  \rho_{\odot} {M_{\odot} R_{\odot}} / {(c_1)_c^4}$
is composed of universal constants, and can be is expressed in terms of the escape velocity for Sun,
$V_{\odot}  = \sqrt{ 2 G M_{ \odot } /R_{\odot} } = 6.16 \times 10^2 \, km / s$, and the speed
of light, $c=2.99 \times 10^8 \, m/s$, as
\begin{equation}\label{b2c}
k_A = \frac{3 \alpha}{4} ( \frac{V_{\odot} }{(c_1)_c} )^4 = 8.633 \times 10^{10} \alpha \rho^{-4/3}
\end{equation}
Finally,
\begin{eqnarray}\label{b2d}
\frac{d m (t)}{dt}  =  k_A
M L \rho
\frac{ m^2 (t) \rho (t) }{ ( \mu^2 (t) + \rho^{2 / 3} (t) )^{3 / 2}}
\end{eqnarray}
where $\rho (t)$ is the normalized dimensionless density of the nebula.

Eqs.~(\ref{em1}), (\ref{em3}) and (\ref{b2d}) complete the system of equations necessary to study the
problem, subject to some assumption about the shape of the spacial distribution profile of nebula
density $\rho( \, s(t) \,)$. 
In our model  
we assume that $\rho (t) = \exp (- s^\nu (t))$ and that
$\nu = 2$.

\end{document}